\documentclass[journal=jpclcd,manuscript=letter]{achemso}
\usepackage{amsmath,amssymb}
\usepackage{float}
\usepackage[version=3]{mhchem} 
\usepackage{color} 

\author{Manish Kumar}
\email{manish.kumar@physics.iitd.ac.in}
\author{Pooja Basera}
\author{Shikha Saini}
\author{Saswata Bhattacharya}
\email{saswata@physics.iitd.ac.in}
\phone{+91-11-2659 1359}
\fax{+91-11-2658 2037}
\affiliation[Indian Institute of Technology Delhi]
{Department of Physics, Indian Institute of Technology Delhi, New Delhi, India}
\title[An \textsf{achemso} demo]
  {Role of Defects in Photocatalytic Water Splitting: Monodoped vs Codoped SrTiO$_3$}
\begin{document}
\begin{abstract}
Using the hybrid density functional theory and \textit{ab initio} atomistic thermodynamics, we report 
monodoping of non-metal (N) or metal (Mn) in SrTiO$_3$ can induce visible light absorption, but none of them are suitable to ameliorate the photocatalytic activity. Therefore, in order to achieve enhanced photocatalytic activity of SrTiO$_3$, we have employed codoped Mn and N simultaneously in SrTiO$_3$ to modulate its electronic properties effectively. In the codoped SrTiO$_3$, the recombination of photogenerated charge carriers is suppressed, and the diffusion and mobility are increased owing to the passivation of discrete localized states. Our results reveal that Mn$_{\textrm{Sr}}\textrm{N}_\textrm{O}$ (codoping of Mn at Sr site and N at O site) is the most promising candidate for enhancing the photocatalytic activity of SrTiO$_3$ under visible light.
\begin{tocentry}
	\begin{figure}[H]
		\includegraphics[width=1.0\columnwidth,clip]{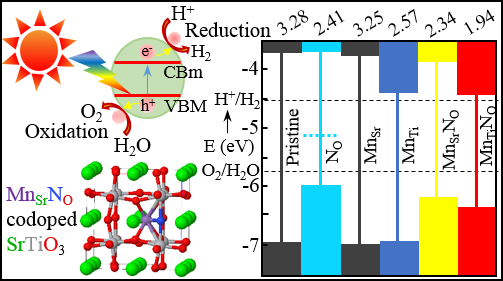}
	\end{figure}
\end{tocentry}
\end{abstract}
\section{Introduction}
SrTiO$_3$ has emerged as one of the most promising energy materials~\cite{doi:10.1021/cr1001645,B921692H,B800489G,doi:10.1021/cr100454n,doi:10.1021/am500157u} for photocatalytic water splitting~\cite{doi:10.1021/acs.jpcc.8b11240, doi:10.1021/j100274a018, AHUJA199699, doi:10.1021/ja2050315, doi:10.1021/acsami.5b11887, NIISHIRO2014187, doi:10.1021/am5051907}. However, owing to its large band gap (3.25 eV)~\cite{doi:10.1021/acsami.7b06025}, its application at a commercial level is delimited. Thus, several works have been endeavored to reduce the band gap of SrTiO$_3$ in order to induce visible light absorption via doping with metals~\cite{doi:10.1021/jp049556p,doi:10.1021/jp300910e,B502147B,C1JM11385B,doi:10.1063/1.4738371}, nonmetals~\cite{LIU201211611,C5CP03005F, WANG2004149,doi:10.1063/1.2403181,ZHANG201369,C2CC33797E} or combination of several elements~\cite{C2TA00450J,doi:10.1021/ja210610h,doi:10.1021/cm503541u,doi:10.1021/la0353125,CHEN2016145}. 
However, the band gap reduction cannot ensure the enhancement in photocatalytic efficiency as it also depends on the location of conduction band minimum (CBm) and valence band maximum (VBM). For transition metal (TM) doped SrTiO$_3$, TM d-states hybridize with those states of the SrTiO$_3$ that contribute to conduction band, and thus, the reduction in band gap occurs by shifting of CBm in downward direction. However, due to this, the reducing power is deteriorated. Also, in general, doping by 3d elements lead to localized states in the gap, which are detrimental to the photocatalysis. 
Therefore, transition metal alone is not suitable for improvement in photocatalytic activity.
On the other hand, nonmetal doped SrTiO$_3$ is found to narrow the band gap by elevating the VBM. 
But in this case also, localized states are appeared deep inside the forbidden region, which can trap the photoexcited charge carriers and accelerate the electron-hole recombination. This in turn degrades the photocatalytic efficiency~\cite{doi:10.1021/jp300910e,doi:10.1021/jp902567j}. This has motivated us for codoping. The codoping with a metal is one of the pre-eminent solutions to passivate such discrete states of dopants in the forbidden region, and form the continuum band~\cite{doi:10.1021/jp902567j,B922399A,PhysRevLett.102.036402,doi:10.1021/jp512948s}. By means of codoping, band edges can be engineered to comply with the needs, i.e., the spectral response expands to the visible region, while retaining the reduction and oxidation power~\cite{doi:10.1021/acs.jpcc.5b06667, LEBAHERS2019212}. A thorough screening of the codopants has been done by calculating the band gap 
(see Table S1 in Supporting Information (SI)). 
From here, we have identified few promising systems (marked as red) and one of them is the N and Mn codoped system. The decrement in band gap is most suitable for the Mn and N codoped system, whereas in rest of the cases, the band gap decrement is either larger or smaller than what is needed for the maximum efficiency in photocatalytic water splitting ($\sim$ 2 eV~\cite{doi:10.1021/cr1002326,C3CP54589J}). Moreover, one of the important factors for choosing Mn is that, it has d-d transition and its d-orbitals' energy facilitates the suitable potentials for water redox reactions. Note that the individual monodopants (i.e. N and Mn) have already been experimentally synthesized~\cite{doi:10.1063/1.2403181,SUN2013176,Liu_2007,TKACH20055061,doi:10.1080/21663831.2013.856815}. However, for the codoping of Mn and N in bulk SrTiO$_3$, any experimental or theoretical reports are hitherto unknown.

In this article, we have, therefore, studied codoped (N-Mn) SrTiO$_3$ for enhancing the photocatalytic efficiency under visible light. To understand the N-Mn codoped case, first we have addressed the respective monodoped (N, Mn) cases and their thermodynamic stability using hybrid density functional theory (DFT)~\cite{PhysRev.136.B864,PhysRev.140.A1133} and \textit{ab initio} atomistic thermodynamics at realistic conditions (temperature $(T)$, partial pressure of oxygen $(p_{\textrm{O}_2})$ and doping)~\cite{PhysRevMaterials.1.071601}. 
Next, to get the insights on synergistic effect of codoping, electronic density of states for pristine, monodoped and codoped SrTiO$_3$ have been compared. In addition, the optical response using single-shot GW~\cite{PhysRev.139.A796,PhysRevLett.55.1418} method is also analyzed. Finally, from the perspective of its usage in photocatalytic water splitting, we have examined the band edges alignment of (un)doped SrTiO$_3$ w.r.t. water redox potential levels.
\section{Computational Methods}
We have carried out the DFT calculations using the Vienna \textit{ab initio} simulation package (VASP)~\cite{KRESSE199615,PhysRevB.59.1758}. The projector-augmented wave (PAW) potentials~\cite{PhysRevB.50.17953} are used to describe the ion-electron interactions in all the elemental constituents, viz. Sr, Ti, Mn, O, and N, that contain ten, four, seven, six, and five valence electrons, respectively. The total energy calculations are performed using hybrid exchange--correlation (xc) functional HSE06~\cite{doi:10.1063/1.2404663} (for validation of xc functionals, see Sec. VI in SI). To introduce defects in SrTiO$_3$, we have used a 40-atom supercell, which is constructed by a $2\,\times\,2\,\times\,2$ repetition of cubic SrTiO$_3$ unit cell (5 atoms). A k-point mesh of $4\times4\times4$ is used, which is generated using Monkhorst-Pack~\cite{PhysRevB.13.5188} scheme. The self consistency loop is converged with a threshold of 0.01 meV energy. The cutoff energy of 600 eV is used for the plane wave basis set. Note that we have performed the spin-polarized calculations because the doped systems contain unpaired electrons. The quasiparticle energy calculations have been carried out using single-shot G$_0$W$_0$ approximation starting from the orbitals obtained using HSE06 xc functional. The polarizability calculations are performed on a grid of 50 frequency points. To make computation feasible, the number of bands is set to 384, which is typically four times the number of occupied orbitals.

\section{Results and discussion}
\subsubsection{Stabilitiy of defects in SrTiO$_3$: \textit{ab initio} atomistic thermodynamics}
On doping SrTiO$_3$ with a nonmetal dopant (e.g. N), the possible defects that could occur are: N$_\textrm{O}$ (N substituted at O position), N$_\textrm{i}$ (N as an interstitial making bond with O), and $(\textrm{N}_2)_\textrm{O}$ split-interstitial
(one N is at interstitial position and another one is substituted the nearby O, making bond with each other)~\cite{scireports2019,doi:10.1063/1.2403181,SUN2013176,Liu_2007}. In the case of metal (e.g Mn) doping, Mn could be substituted either at Ti (Mn$_\textrm{Ti}$) or Sr (Mn$_\textrm{Sr}$) site, or it could also be present as an interstitial (Mn$_\textrm{i}$) in SrTiO$_3$~\cite{TKACH20055061,doi:10.1080/21663831.2013.856815}. Note that these defects are not stable in neutral form because of the uncompensated charge. Therefore, we have calculated the stability of charged defects in addition to neutral defects with charge states $q$ ($-2$, $-1$, $0$, $+1$, $+2$)~\cite{PhysRevMaterials.1.071601,PhysRevB.94.094305,scireports2019}. 

To analyze the thermodynamic stability of the defected configuration w.r.t. pristine SrTiO$_3$, we have calculated the formation energy by means of \textit{ab initio} atomistic thermodynamics~\cite{Bhattacharya_2014,doi:10.1021/acs.jpclett.5b01435,PhysRevB.94.094305,ekta-jpcc, PhysRevMaterials.1.071601}. 
For a X-related defect with charge state $q$, the formation energy ($\textrm{E}_\textrm{f}(\textrm{X}^{q}$)) is calculated as follow~\cite{scireports2019,PhysRevMaterials.1.071601,PhysRevB.94.094305}:
\begin{equation}\begin{split}
\textrm{E}_\textrm{f}(\textrm{X}^{q}) &= \textrm{E}_\textrm{tot}(\textrm{X}^{q}) - \textrm{E}_\textrm{tot}(\textrm{pristine}^{0}) - \sum_{i} {n}_i\mu_{i}\\
&\quad+ {q}(\mu_\textrm{e} + \textrm{VBM} + \Delta\textrm{V})\textrm{,}
\end{split}\end{equation}
where, $\textrm{E}_\textrm{tot}(\textrm{X}^{q})$ and $\textrm{E}_\textrm{tot}(\textrm{pristine}^{0})$ are the total DFT energies with defect (at charge $q$) and pristine neutral respectively. ${n}_i$ is the number of atoms $i$ added (positive) or removed (negative) from the system and $\mu_{i}$ is the corresponding chemical potential. $\mu_{i}$ is referenced from the total DFT energy $(\textrm{E}_\textrm{tot}(i))$ of species $i$, i.e., $\mu_{i} = \Delta\mu_{i} + \textrm{E}_\textrm{tot}(i)$. The chemical potentials, $\Delta\mu_{i}$s have been chosen carefully to reflect the appropriate environmental growth conditions (see Sec. II in SI). $\mu_\textrm{e}$ is the chemical potential of electron varied from VBM to CBm of the pristine system and $\Delta\textrm{V}$ accounts for the core level alignment of the defected system w.r.t. the pristine neutral system.
\begin{figure}[H]
	\centering
	\includegraphics[width=0.6\textwidth]{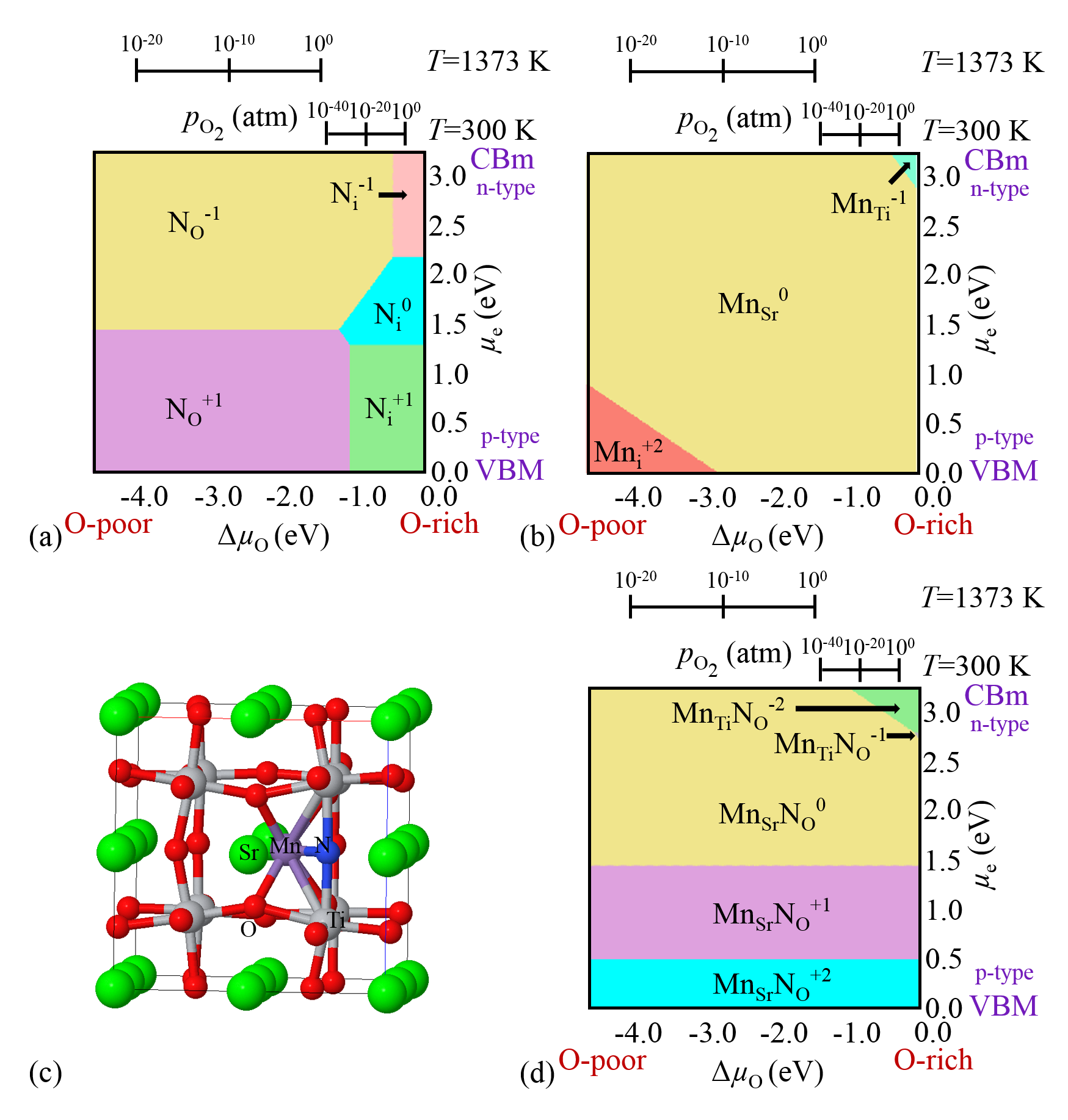} 
	\caption{
	2D projection of the 3D phase diagram that manifests the stable phases of (a) N-related, (b) Mn-related and (d) (N-Mn)-related charged defects having minimum formation energy as a function of $\mu_\textrm{e}$ and $\Delta\mu_\textrm{O}$. Here, on x-axis, $\Delta\mu_\textrm{O}$ is varied according to $T$ and $p_{\textrm{O}_2}$, and on y-axis, $\mu_\textrm{e}$ is varied from VBM to CBm of pristine SrTiO$_3$. Colored regions show the most stable phases having minimum formation energy at a given environmental condition. Top axes are showing the pressure $(p_{\textrm{O}_2})$ range at two temperatures: $T$=300 K and 1373 K. (c) Ball and stick model of optimized structure of Mn$_\textrm{Sr}$N$_\textrm{O}$ defect configuration.}
	\label{fig_N-rel}
\end{figure}
The optimized structures of all the doped and pristine SrTiO$_3$ supercell are shown in Figure S1. SrTiO$_3$ has a cubic structure space group \textit{Pm}$\bar{3}$\textit{m} at room temperature. On doping N in pristine, N-related defects, viz. N$_\textrm{O}$, N$_\textrm{i}$, and $(\textrm{N}_2)_\textrm{O}$ could form. N$_\textrm{O}$ shows negligible distortion, whereas N$_\textrm{i}$ and $(\textrm{N}_2)_\textrm{O}$ show more distortion in the lattice.
We can sum up about the stability of all the three configurations of N-related defects at different environmental conditions by observing the 3D phase diagram as shown in Figure~\ref{fig_N-rel}a. Here, on x-axis, $\Delta\mu_\textrm{O}$ is varied from O-poor to O-rich condition in accordance with $T$ and $p_{\textrm{O}_2}$. On y-axis, we have scanned the entire forbidden region by means of $\mu_\textrm{e}$, which is referenced from VBM of pristine SrTiO$_3$. On z-axis, we have shown the most stable phases having minimum formation energy at a given environmental condition using the colored surfaces. The charge states $+1$ and $-1$ are energetically stable in case of N$_\textrm{O}$ near VBM and CBm, respectively. N$_\textrm{i}$ is energetically stable in charge states $+1$, $0$, and $-1$. The positive charge states are more favorable for smaller value of $\mu_\textrm{e}$, i.e., near VBM (p-type), whereas negative charge states are stable near CBm (n-type) for larger value of $\mu_\textrm{e}$. Since O-poor and O-rich conditions also correspond to lesser and more content of O, respectively, therefore N$_\textrm{O}$ is difficult/easier to form in O-rich/O-poor condition. N$_\textrm{O}$ is stable with charge state $-1$ near CBm as it has one electron less than the O atom. The thermodynamic transition level (+/$-$) lies in between the VBM and CBm indicating that N$_\textrm{O}$ acts both as a deep donor/acceptor depending on the nature of doping (i.e. p-type or n-type). From Figure~\ref{fig_N-rel}a, we can easily see that N$_\textrm{O}$ is the predominant defect in N doped SrTiO$_3$ for a wide range of environmental conditions including the experimental growth condition ($T = 1373 \textrm{ K}$, $p_{\textrm{O}_2} = 1$ atm~\cite{WANG2004149}), whereas N$_\textrm{i}$ is only favorable in O-rich condition (in accordance with 2D phase diagram (see Figure S2)).

On doping Mn in SrTiO$_3$, the structures that could form are Mn$_\textrm{Sr}$, Mn$_\textrm{Ti}$, and Mn$_\textrm{i}$.
In case of Mn$_\textrm{Sr}$, only neutral defect is stable, which signifies that Mn exists in Mn$^{2+}$ oxidation state when substituted at Sr (Sr$^{2+}$ oxidation state) site in SrTiO$_3$ (see Figure~\ref{fig_N-rel}b). Mn$_\textrm{Ti}$ is stable in $-1$ charge state indicating that in addition to Mn$^{4+}$ oxidation state, Mn$^{3+}$ oxidation state could also exist, though unlikely, when Mn is substituted at Ti (Ti$^{4+}$ oxidation state) site. Mn$_\textrm{i}$ with $+2$ charge state is stable in p-type SrTiO$_3$ under O-poor condition, while Mn$_\textrm{Ti}$ with -1 charge state is stable in n-type SrTiO$_3$ under O-rich condition as shown in the 3D phase diagram (see Figure~\ref{fig_N-rel}b). Neutral Mn$_\textrm{Sr}$ is the prominent defect under all the three environmental conditions. The formation energy for Mn doped SrTiO$_3$ in all oxygen environmental conditions is small, particularly in O-intermediate condition (see Figure S3), which implies that it is easier to dope Mn in SrTiO$_3$.

From the above analysis, we conclude that in the case of monodoped SrTiO$_3$, substitutional doping is the most stable for a wider region of the environmental conditions including the experimental growth conditions. Therefore, 
we have considered only the substitutional position for codoping of Mn and N in SrTiO$_3$. 
The formation energy of Mn$_\textrm{Sr}$N$_\textrm{O}$ ($\textrm{E}_\textrm{f}(\textrm{Mn}_\textrm{Sr}\textrm{N}_\textrm{O}^{q})$) is calculated as follow:
\begin{equation}\begin{split}
\textrm{E}_\textrm{f}(\textrm{Mn}_\textrm{Sr}\textrm{N}_\textrm{O}^{q}) &= \textrm{E}_\textrm{tot}(\textrm{Mn}_\textrm{Sr}\textrm{N}_\textrm{O}^{q}) - \textrm{E}_\textrm{tot}(\textrm{SrTiO}_3^{0}) + \mu_\textrm{O} - \mu_\textrm{N}\\ &\quad+ \mu_\textrm{Sr} - \mu_\textrm{Mn} + {q}(\mu_\textrm{e} + \textrm{VBM} + \Delta\textrm{V})\textrm{,}
\end{split}\end{equation}
where $\textrm{E}_\textrm{tot}(\textrm{Mn}_\textrm{Sr}\textrm{N}_\textrm{O}^{q})$ and $\textrm{E}_\textrm{tot}(\textrm{SrTiO}_3^{0})$ are the DFT energies of the codoped system (Mn at Sr and N at O) with charge $q$ and the pristine neutral SrTiO$_3$, respectively. $\mu_\textrm{O}$ and $\mu_\textrm{N}$ are the chemical potential of oxygen and nitrogen atom, referenced from the total DFT energy with addition of zero point energy of O$_2$ and N$_2$ molecules, respectively, i.e., $\mu_\textrm{O} = \Delta\mu_\textrm{O} + \frac{1}{2}\left(\textrm{E}_\textrm{tot}(\textrm{O}_2) + \frac{\textrm{h}\nu_\textrm{OO}}{2}\right)$ and $\mu_\textrm{N} = \Delta\mu_\textrm{N} + \frac{1}{2}\left(\textrm{E}_\textrm{tot}(\textrm{N}_2) + \frac{\textrm{h}\nu_\textrm{NN}}{2}\right)$. In the latter terms, $\nu_\textrm{OO}$ and $\nu_\textrm{NN}$ are the O-O and N-N stretching frequencies, respectively. The chemical potential $\mu_\textrm{X} = \Delta\mu_\textrm{X} + \textrm{E}_\textrm{tot}(\textrm{X})$ (where, X = Mn, Sr and Ti). The chemical potential $\Delta\mu_\textrm{X}$ (X = O, N, Mn, Sr, and Ti) are chosen carefully (see Sec. II in SI). Figure~\ref{fig_N-rel}d shows the 3D phase diagram for the stability of codoped systems.  
Mn$_\textrm{Sr}$N$_\textrm{O}$ is the predominant defect in all the environmental growth conditions and is stable in +2, +1, and neutral charge states. This will act as a donor in p-type SrTiO$_3$. Whereas, Mn$_\textrm{Ti}$N$_\textrm{O}$ is stable only for a smaller region in extreme O-rich/Ti-poor condition with charge states -1 and -2 near CBm (n-type), i.e. it will act as an acceptor (see Figure~\ref{fig_N-rel}d). To further confirm that the formation of defect pair (N-Mn) in SrTiO$_3$ is stable,
the binding energy ($\textrm{E}_\textrm{b}$) of the defect pairs (N-Mn) in SrTiO$_3$ has also been checked~\cite{PhysRevB.74.081201,PhysRevLett.102.036402}.
A more negative value of $\textrm{E}_\textrm{b}$ indicates that defect pair is more stable when both the dopants are present in the sample. $\textrm{E}_\textrm{b}$ for Mn$_\textrm{Sr}$N$_\textrm{O}$ and Mn$_\textrm{Ti}$N$_\textrm{O}$ pairs are $-1.46$ and $-0.33$ eV, respectively. These values indicate that defect pairs are more stable than the isolated impurities in SrTiO$_3$ supercell. Also, Mn$_\textrm{Sr}$N$_\textrm{O}$ is more stable configuration than Mn$_\textrm{Ti}$N$_\textrm{O}$ since Mn$_\textrm{Sr}$N$_\textrm{O}$ has higher (more negative) binding energy than Mn$_\textrm{Ti}$N$_\textrm{O}$. In the codoped system, Mn acts as a donor, whereas N acts as an acceptor. The charge transfer takes place from donor to acceptor and strong Coulomb interaction arises between positively charged donor and negatively charged acceptor. Hence, the defect pair is stable. The extra stability in Mn$_\textrm{Sr}$N$_\textrm{O}$ is due to the shift of Mn away from the Sr center towards N as shown in Figure~\ref{fig_N-rel}c and making strong bonds with its neighbor atoms.

\subsubsection{Electronic structure analysis}
\begin{figure}
	\centering
	\includegraphics[width=0.6\textwidth]{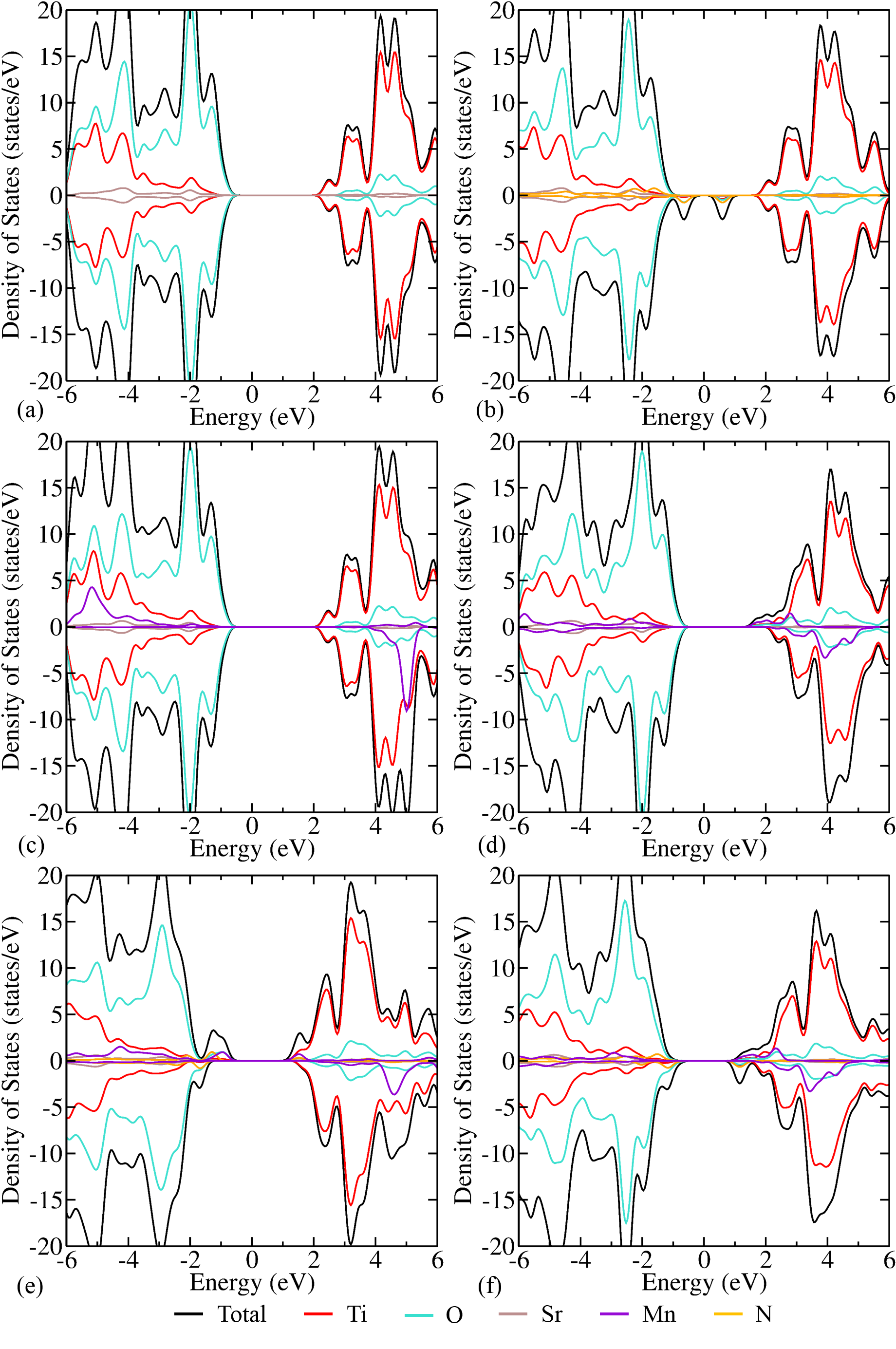}
	\caption{Electronic density of states for the supercell of (a) pristine SrTiO$_3$, (b) N$_\textrm{O}$, (c) Mn$_\textrm{Sr}$, (d) Mn$_\textrm{Ti}$, (e) Mn$_\textrm{Sr}\textrm{N}_\textrm{O}$ and (f) Mn$_\textrm{Ti}\textrm{N}_\textrm{O}$ type defect.}
	\label{fig_pris_N_DOS}
\end{figure}

To get more insights about the effect of dopants in SrTiO$_3$, we have calculated atom projected density of states (pDOS). The DOS plots for pristine and monodoped SrTiO$_3$ are shown in Figure~\ref{fig_pris_N_DOS}(a-d).
In pristine, the O 2p orbitals contribute to VBM, and Ti 3d orbitals contribute to CBm with a wide band gap of 3.28 eV. The DOS of pristine is symmetric w.r.t. spin alignments (i.e. spin up or down), whereas in case of N$_\textrm{O}$, the DOS is asymmetrical due to devoid of an electron in comparison to pristine SrTiO$_3$ (see Figure~\ref{fig_pris_N_DOS}a and~\ref{fig_pris_N_DOS}b). In the latter case, some occupied states are appeared above pristine-VBM and some unoccupied discrete states can also be seen deep inside the forbidden region (since the N 2p orbitals have higher energy than the O 2p orbitals) (see Figure~\ref{fig_pris_N_DOS}b). This leads to reduction in the band gap. 
However, these midgap states increase the recombination rate and decrease the charge mobility that lead to degradation in the photocatalytic activity.

We have considered two sites for the substitution of Mn, viz. Sr and Ti sites. In case of monodoping of Mn at Sr site, the band gap (3.25 eV) is not getting reduced and thus, cannot induce visible light absorption (see Figure~\ref{fig_pris_N_DOS}c). The occupied and unoccupied states of Mn orbitals are appeared deep inside the valence and conduction band, respectively, indicating that Mn$_\textrm{Sr}$ is very stable. However, in case of Ti site substitution, we get interesting feature in the DOS (see Figure~\ref{fig_pris_N_DOS}d). The localized states bring down the CBm. Hence, the band gap is reduced to 2.57 eV, resulting in the visible light absorption. However, due to shift of CBm in downward direction, its reduction power is degraded. Therefore, it can not be a potential candidate for H$_2$ production from water splitting.

In case of codoping, the substitution of Mn at both sites, Sr and Ti in addition to N$_\textrm{O}$, helps in passivating the localized mid gap states (introduced by N substitution) and form continuum states as shown in Figure~\ref{fig_pris_N_DOS}e and~\ref{fig_pris_N_DOS}f.
The passivation of states is concomitant with the hybridization of O and N orbitals, and Mn and O orbitals in Mn$_\textrm{Sr}\textrm{N}_\textrm{O}$ defect configuration as shown in Figure~\ref{fig_pris_N_DOS}e (near VBM). However, in the case of Mn$_\textrm{Ti}\textrm{N}_\textrm{O}$, Mn states arise only near CBm as shown in Figure~\ref{fig_pris_N_DOS}f. The recombination of photogenerated charge carriers is suppressed, and the diffusion and mobility are increased owing to the passivation of discrete localized states. The band gaps of Mn$_\textrm{Sr}\textrm{N}_\textrm{O}$ and Mn$_\textrm{Ti}\textrm{N}_\textrm{O}$ are 2.34 and 1.94 eV, respectively, which are the desirable one for visible light absorption. In case of Mn$_\textrm{Ti}\textrm{N}_\textrm{O}$, CBm is shifted downward by a large amount and hence, adversely affect the reduction power for hydrogen generation. However, in case of Mn$_\textrm{Sr}\textrm{N}_\textrm{O}$, this downward shift is very small. Consequently, the codoping of Mn at Sr site and N at O site is favorable for overall photocatalytic water splitting. Also, from Figure~\ref{fig_N-rel}c, we can see a relatively large distortion in case of Mn$_\textrm{Sr}\textrm{N}_\textrm{O}$ codoping, which builds up the internal field, that is helpful for photogenerated charge carriers separation and thus, enhances the photocatalytic efficiency. Therefore, Mn$_\textrm{Sr}\textrm{N}_\textrm{O}$ codoping in SrTiO$_3$ is the promising candidate to enhance the photocatalytic efficiency and generate hydrogen from water splitting.

\subsubsection{Optical properties}
\begin{figure}
	\centering
	\includegraphics[width=0.6\textwidth]{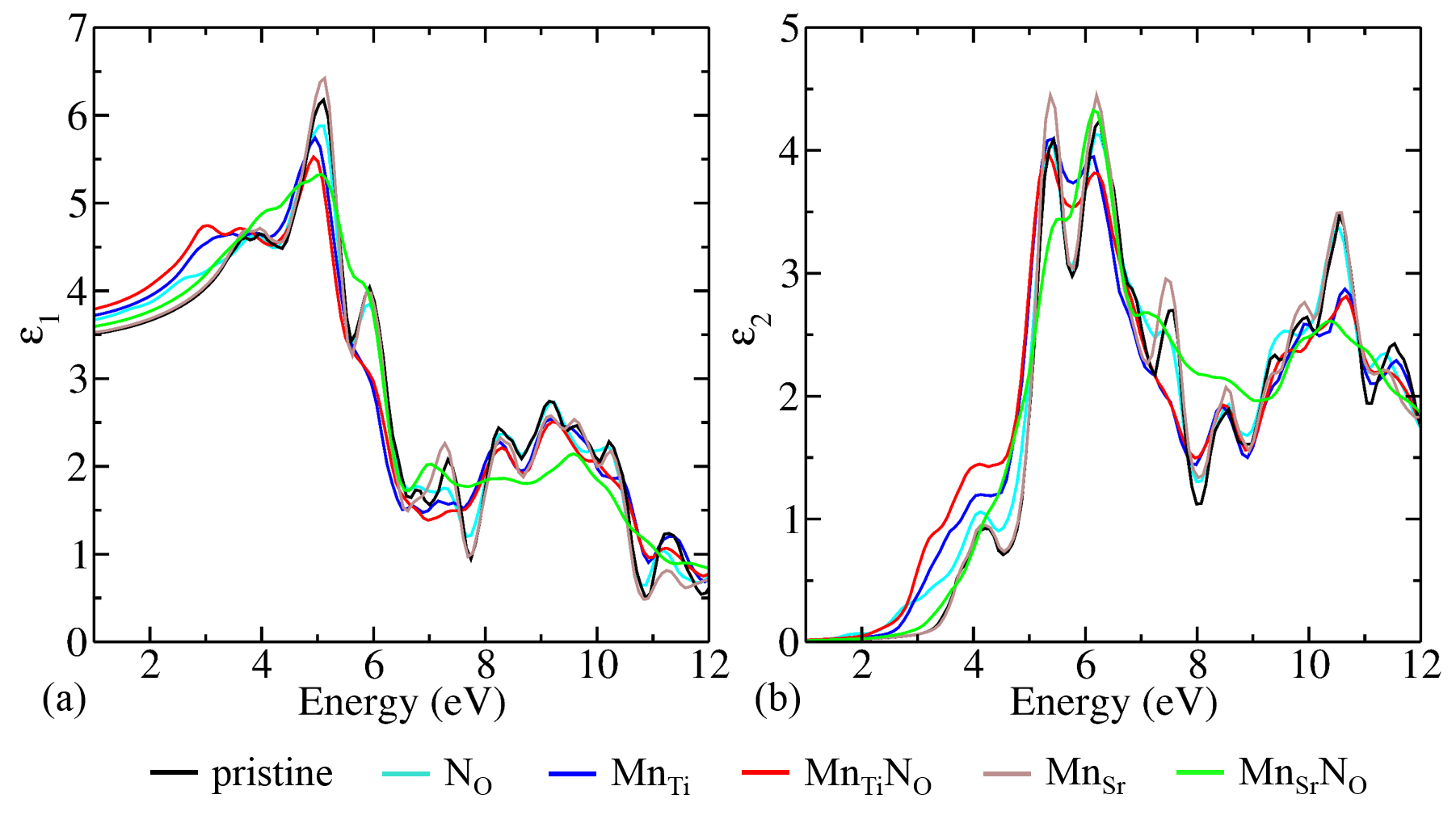}
	\caption{Spatially average (a) real $(\epsilon_1)$ and (b) imaginary $(\epsilon_2)$ part of the dielectric function obtained by G$_0$W$_0$@HSE06 for the pristine, monodoped and codoped SrTiO$_3$.}
	\label{fig_optical}
\end{figure}

To determine the optical spectra, we have computed the frequency dependent complex dielectric function $\epsilon(\omega) = \epsilon_1(\omega) + i\epsilon_2(\omega)$ using the G$_0$W$_0$@HSE06 approach (the results with HSE06 are shown in Figure S5 and S6). The real ($\epsilon_1$) and imaginary ($\epsilon_2$) part of dielectric function have shown in Figure~\ref{fig_optical}a and~\ref{fig_optical}b, respectively.
The static $(\omega = 0)$ real part of $\epsilon(\omega)$ for pristine SrTiO$_3$ is found to be 3.46, which is well in agreement with previous findings~\cite{PhysRevMaterials.2.024601}. On doping, its value is increased (see Figure~\ref{fig_optical}a).
The imaginary part of dielectric function represents the optical absorption (see Figure~\ref{fig_optical}b). We have found the first peak at 4.20 eV for pristine SrTiO$_3$ [experimental value is 4.7 eV~\cite{PhysRev.140.A651}].  
Note that the peaks are shifted towards lower energy region, except for Mn$_\textrm{Sr}$. Even if the exact numbers may differ from the experimental values, atleast from the trends it's clear that the band gap is getting reduced on doping in SrTiO$_3$. Also, we could note that the onset of the absorption edge is shifting towards the lower region. Hence, the optical response is shifted towards the visible region. 
The spectra of Mn$_\textrm{Sr}$ coincides with the pristine supercell of SrTiO$_3$ because there is no reduction in band gap, while in rest of the cases, band gap is reduced (see Figure~\ref{fig_optical}b).

\subsubsection{Band edge alignment}
\begin{figure}
	\centering
	\includegraphics[width=0.6\textwidth]{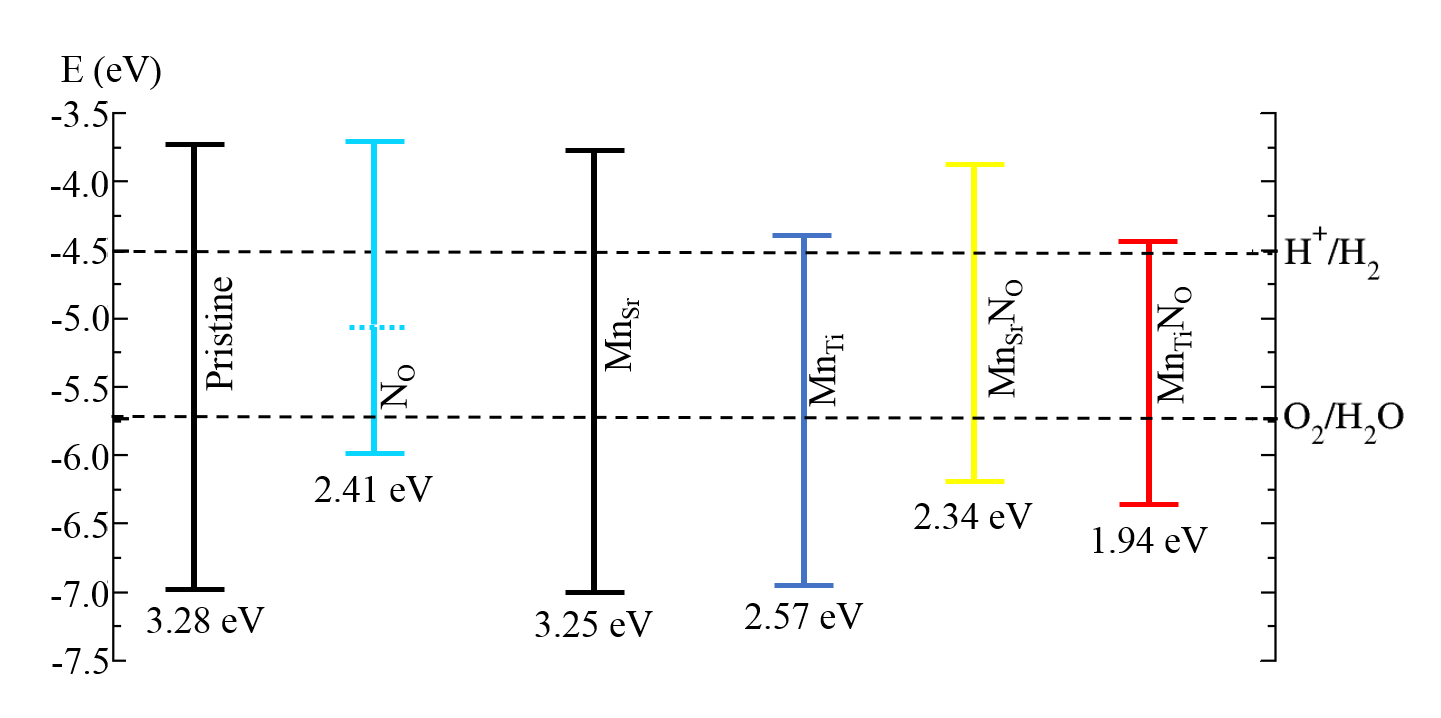}
	\caption{Band edge alignment of (un)doped SrTiO$_3$ w.r.t. water redox potential levels (H$^+$/H$_2$, O$_2$/H$_2$O).}
	\label{fig_band_edge_align}
\end{figure}

Note that only reduction in band gap can not assure the hydrogen generation via photocatalytic water splitting. The band edges (VBM and CBm) should have appropriate position. For water splitting, the CBm must lie above the water reduction potential level (H$^+$/H$_2$) and VBM must be positioned below water oxidation potential level (O$_2$/H$_2$O). Firstly, we have aligned the band edges of undoped SrTiO$_3$ w.r.t. water redox potential levels. The CBm lies 0.8 eV above the water reduction potential (H$^+$/H$_2$) and VBM lies 1.25 eV below water oxidation potential~\cite{band_edge}. We see that the position of CBm and VBM of the pristine SrTiO$_3$ is consistent with previous findings~\cite{C3CP54589J}. Thereafter, we align the band edges of doped SrTiO$_3$ by observing the shift in energy of the VBM and CBm w.r.t. undoped SrTiO$_3$. From Figure~\ref{fig_band_edge_align}, we have found that in the case of N$_\textrm{O}$, the VBM is shifted upwards and the CBm is not disturbed.
However, some localized states are present deep in the forbidden region, which degrade the photocatalytic efficiency. Hence, N$_\textrm{O}$ is not the promising one for water splitting. In case of Mn$_\textrm{Ti}$ and Mn$_\textrm{Ti}\textrm{N}_\textrm{O}$, the CBm is shifted downward by a large amount and hence, their reduction power is very low and could not be utilized for hydrogen generation from water. On substituting Mn at Sr site,  the band gap is not getting reduced and thus, not inducing the visible light response. In the case of Mn$_\textrm{Sr}\textrm{N}_\textrm{O}$, CBm is shifted by small amount and thus, retaining intact the reduction power. It has a desirable band gap of 2.34 eV and also, does not contain any localized midgap states. In view of this, from the applicability in photocatalytic water splitting, only Mn$_\textrm{Sr}\textrm{N}_\textrm{O}$ is the most desirable one. Hence, these theoretical studies help in further future investigations to engineer a device that will be environment friendly.
\subsubsection{Effective mass for pristine and Mn$_\textrm{Sr}$N$_\textrm{O}$ codoped SrTiO$_3$}
We have also determined the effective mass of electrons and holes for the pristine and Mn$_\textrm{Sr}$N$_\textrm{O}$ codoped SrTiO$_3$ configurations using HSE06 functional (see Table~\ref{tbl}). These have been obtained from the curvature of band edges by calculating inverse of second derivative of band energies with respect to k at band edges. These values of effective masses for pristine system (except for heavy-hole) are validated with previous studies~\cite{PhysRevB.84.201304,doi:10.1063/1.1570922,C8TA09022J}.

\begin{figure}
	\centering
	\includegraphics[width=0.6\textwidth]{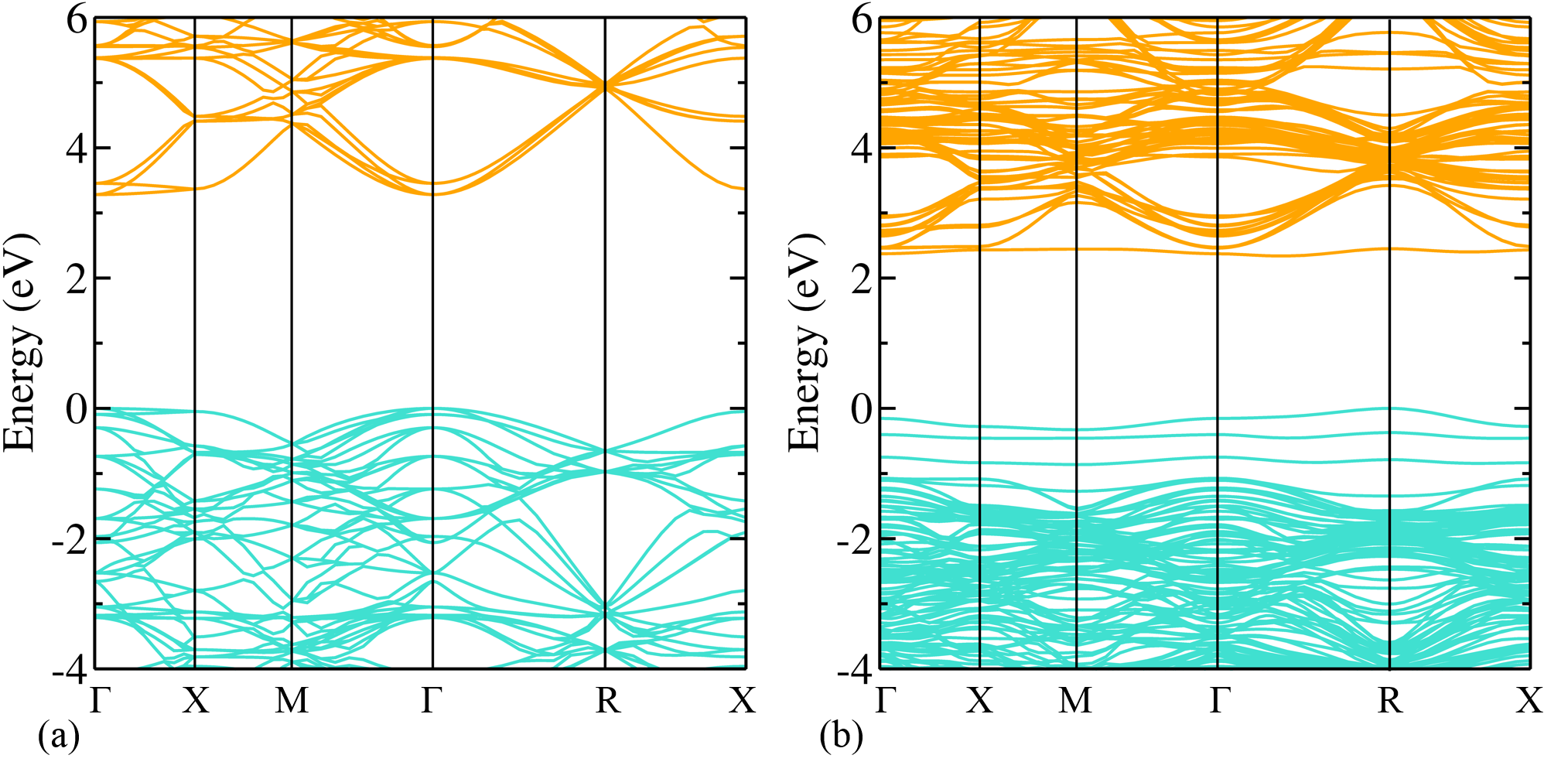}
	\caption{Band structure calculated using hybrid (HSE06) functional of (a) pristine SrTiO$_3$ and (b) Mn$_\textrm{Sr}$N$_\textrm{O}$ codoped SrTiO$_3$.}
	\label{fig_band_str}
\end{figure}

\begin{table}
	\caption{Effective masses (in terms of free-electron mass m$_\textrm{e}$) at the band edge for pristine and Mn$_\textrm{Sr}$N$_\textrm{O}$ codoped SrTiO$_3$.  The masses $\textrm{m}_\textrm{he}, \textrm{m}_\textrm{le},\textrm{m}_\textrm{hh}, \textrm{and}\;\textrm{m}_\textrm{lh}$ correspond to heavy-electron, light-electron, heavy-hole, and light-hole band, respectively.}
	\label{tbl}
	\begin{tabular}{lllll}
		\hline
		Configuration & $\;\;\:\textrm{m}_\textrm{he}$ & $\textrm{m}_\textrm{le}$ & $\;\;\:\textrm{m}_\textrm{hh}$ & $\;\;\:\textrm{m}_\textrm{lh}$\\
		\hline
		pristine & $\;\;\:5.18$ & $0.38$  & $-10.36$ & $-0.74$ \\
		Mn$_\textrm{Sr}$N$_\textrm{O}$ & $\;\;\:5.09$ & $$ & $-2.58$ & $$\\
		\hline
	\end{tabular}
\end{table}
From Figure~\ref{fig_band_str}a, we can see that the pristine has three-fold degeneracy at CBm (at $\Gamma$ k-point). This degeneracy is lifted as one moves away from $\Gamma$ k-point in the direction of X, M, or, R k-point. The effective mass of electron/hole due to heavy-, light-electron/hole and spin split-off band is obtained along high symmetry $\Gamma$-X path for pristine. The effective mass of electron/hole corresponding to spin split-off band is found to be same as that for light-electron/hole band.
After the validation for pristine system, we have calculated the effective mass for Mn$_\textrm{Sr}$N$_\textrm{O}$. From Figure~\ref{fig_band_str}b, we can see that there is no degeneracy of band at CBm and VBM. The defect states constitute the CBm and VBM, which are contributed by the codoping of Mn at Sr-site and N at O-site. The flat bands (smaller curvature) indicate that the defect is localized in the system. The CBm for Mn$_\textrm{Sr}$N$_\textrm{O}$ lies at [0.2 0.2 0.2] k-point and VBM lies at R k-point. The effective mass of electron is 5.09m$_\textrm{e}$ and 3.04m$_\textrm{e}$ along [0.2 0.2 0.2]-$\Gamma$ and [0.2 0.2 0.2]-X directions, respectively. This implies the anisotropy in effective mass. The effective mass of hole is -2.58m$_\textrm{e}$ and -1.53m$_\textrm{e}$ along R-$\Gamma$ and R-X directions, respectively. The comparable values of effective mass of electron for codoped and pristine systems indicate that the  mobility will not be affected much due to the shallower defect level. These effective mass studies further assist future experimental and theoretical investigations to tailor the transport properties of the system.
\section{Conclusions}
In summary, we have systematically studied the thermodynamic stability of different types of dopants and codopants in SrTiO$_3$ using hybrid DFT and \textit{ab initio} thermodynamics. Our results indicate that Mn$_\textrm{Sr}\textrm{N}_\textrm{O}$ is the stable defect configuration under equilibrium growth conditions. We have found that 
despite monodopant-related defects help in reducing the band gap, they generate localized states deep inside the band gap. These states act as recombination centers, which in turn reduce the photocatalytic efficiency.
Thus, monodoping of both N or Mn can induce visible light absorption, but none of them are suitable for photocatalytic activity. The codoping 
reduces the band gap to ideal visible region as well as passivates the localized states to form the continuous band with suitable band edge positions. We have observed that the calculated effective mass of electron for codoped system is found to be similar to that of pristine system. This indicates that electron mobility is not reduced much due to the shallower defect level. We find that 
Mn$_{\textrm{Sr}}\textrm{N}_\textrm{O}$ codoped SrTiO$_3$ could be a potential candidate for producing hydrogen via photocatalytic water splitting.
\begin{acknowledgement}
MK acknowledges CSIR, India, for the junior research fellowship [grant no. 09/086(1292)/2017-EMR-I]. PB acknowledges UGC, India, for the senior research fellowship [grant no. 20/12/2015 (ii) EUV]. SS acknowledges CSIR, India, for the senior research fellowship [grant no. 09/086
(1231)/2015-EMR-I]. SB acknowledges the financial support from YSS-SERB research grant, DST, India (grant no. YSS/2015/001209). We acknowledge the High Performance Computing (HPC) facility at IIT Delhi for computational resources.
\end{acknowledgement}
\begin{suppinfo}
Details of choice of the codopants and chemical potentials, validation of exchange-correlation functional, and optical response using hybrid functional could be found in the supporting information file.
\end{suppinfo}
\bibliography{achemso-demo}

\providecommand{\latin}[1]{#1}
\makeatletter
\providecommand{\doi}
  {\begingroup\let\do\@makeother\dospecials
  \catcode`\{=1 \catcode`\}=2 \doi@aux}
\providecommand{\doi@aux}[1]{\endgroup\texttt{#1}}
\makeatother
\providecommand*\mcitethebibliography{\thebibliography}
\csname @ifundefined\endcsname{endmcitethebibliography}
  {\let\endmcitethebibliography\endthebibliography}{}
\begin{mcitethebibliography}{64}
\providecommand*\natexlab[1]{#1}
\providecommand*\mciteSetBstSublistMode[1]{}
\providecommand*\mciteSetBstMaxWidthForm[2]{}
\providecommand*\mciteBstWouldAddEndPuncttrue
  {\def\EndOfBibitem{\unskip.}}
\providecommand*\mciteBstWouldAddEndPunctfalse
  {\let\EndOfBibitem\relax}
\providecommand*\mciteSetBstMidEndSepPunct[3]{}
\providecommand*\mciteSetBstSublistLabelBeginEnd[3]{}
\providecommand*\EndOfBibitem{}
\mciteSetBstSublistMode{f}
\mciteSetBstMaxWidthForm{subitem}{(\alph{mcitesubitemcount})}
\mciteSetBstSublistLabelBeginEnd
  {\mcitemaxwidthsubitemform\space}
  {\relax}
  {\relax}

\bibitem[Chen \latin{et~al.}(2010)Chen, Shen, Guo, and
  Mao]{doi:10.1021/cr1001645}
Chen,~X.; Shen,~S.; Guo,~L.; Mao,~S.~S. Semiconductor-based photocatalytic
  hydrogen generation. \emph{Chemical Reviews} \textbf{2010}, \emph{110},
  6503--6570\relax
\mciteBstWouldAddEndPuncttrue
\mciteSetBstMidEndSepPunct{\mcitedefaultmidpunct}
{\mcitedefaultendpunct}{\mcitedefaultseppunct}\relax
\EndOfBibitem
\bibitem[Chen \latin{et~al.}(2010)Chen, Ma, and Zhao]{B921692H}
Chen,~C.; Ma,~W.; Zhao,~J. Semiconductor-mediated photodegradation of
  pollutants under visible-light irradiation. \emph{Chem. Soc. Rev.}
  \textbf{2010}, \emph{39}, 4206--4219\relax
\mciteBstWouldAddEndPuncttrue
\mciteSetBstMidEndSepPunct{\mcitedefaultmidpunct}
{\mcitedefaultendpunct}{\mcitedefaultseppunct}\relax
\EndOfBibitem
\bibitem[Kudo and Miseki(2009)Kudo, and Miseki]{B800489G}
Kudo,~A.; Miseki,~Y. Heterogeneous photocatalyst materials for water splitting.
  \emph{Chem. Soc. Rev.} \textbf{2009}, \emph{38}, 253--278\relax
\mciteBstWouldAddEndPuncttrue
\mciteSetBstMidEndSepPunct{\mcitedefaultmidpunct}
{\mcitedefaultendpunct}{\mcitedefaultseppunct}\relax
\EndOfBibitem
\bibitem[Kubacka \latin{et~al.}(2012)Kubacka, Fern\'andez-Garc\'ia, and
  Col\'on]{doi:10.1021/cr100454n}
Kubacka,~A.; Fern\'andez-Garc\'ia,~M.; Col\'on,~G. Advanced nanoarchitectures
  for solar photocatalytic applications. \emph{Chemical Reviews} \textbf{2012},
  \emph{112}, 1555--1614\relax
\mciteBstWouldAddEndPuncttrue
\mciteSetBstMidEndSepPunct{\mcitedefaultmidpunct}
{\mcitedefaultendpunct}{\mcitedefaultseppunct}\relax
\EndOfBibitem
\bibitem[Manikandan \latin{et~al.}(2014)Manikandan, Tanabe, Li, Ueda, Ramesh,
  Kodiyath, Wang, Hara, Dakshanamoorthy, Ishihara, Ariga, Ye, Umezawa, and
  Abe]{doi:10.1021/am500157u}
Manikandan,~M.; Tanabe,~T.; Li,~P.; Ueda,~S.; Ramesh,~G.~V.; Kodiyath,~R.;
  Wang,~J.; Hara,~T.; Dakshanamoorthy,~A.; Ishihara,~S. \latin{et~al.}
  Photocatalytic water splitting under visible light by mixed-valence
  Sn${}_{3}$O${}_{4}$. \emph{ACS Applied Materials \& Interfaces}
  \textbf{2014}, \emph{6}, 3790--3793\relax
\mciteBstWouldAddEndPuncttrue
\mciteSetBstMidEndSepPunct{\mcitedefaultmidpunct}
{\mcitedefaultendpunct}{\mcitedefaultseppunct}\relax
\EndOfBibitem
\bibitem[Zhao \latin{et~al.}(2019)Zhao, Liang, Ma, Huang, and
  Dai]{doi:10.1021/acs.jpcc.8b11240}
Zhao,~P.; Liang,~Y.; Ma,~Y.; Huang,~B.; Dai,~Y. Janus chromium dichalcogenide
  monolayers with low carrier recombination for photocatalytic overall
  water-splitting under infrared light. \emph{The Journal of Physical Chemistry
  C} \textbf{2019}, \emph{123}, 4186--4192\relax
\mciteBstWouldAddEndPuncttrue
\mciteSetBstMidEndSepPunct{\mcitedefaultmidpunct}
{\mcitedefaultendpunct}{\mcitedefaultseppunct}\relax
\EndOfBibitem
\bibitem[Domen \latin{et~al.}(1986)Domen, Kudo, Onishi, Kosugi, and
  Kuroda]{doi:10.1021/j100274a018}
Domen,~K.; Kudo,~A.; Onishi,~T.; Kosugi,~N.; Kuroda,~H. Photocatalytic
  decomposition of water into hydrogen and oxygen over nickel(II)
  oxide-strontium titanate (SrTiO${}_{3}$) powder. 1. Structure of the
  catalysts. \emph{The Journal of Physical Chemistry} \textbf{1986}, \emph{90},
  292--295\relax
\mciteBstWouldAddEndPuncttrue
\mciteSetBstMidEndSepPunct{\mcitedefaultmidpunct}
{\mcitedefaultendpunct}{\mcitedefaultseppunct}\relax
\EndOfBibitem
\bibitem[Ahuja and Kutty(1996)Ahuja, and Kutty]{AHUJA199699}
Ahuja,~S.; Kutty,~T. Nanoparticles of SrTiO${}_{3}$ prepared by gel to
  crystallite conversion and their photocatalytic activity in the
  mineralization of phenol. \emph{Journal of Photochemistry and Photobiology A:
  Chemistry} \textbf{1996}, \emph{97}, 99--107\relax
\mciteBstWouldAddEndPuncttrue
\mciteSetBstMidEndSepPunct{\mcitedefaultmidpunct}
{\mcitedefaultendpunct}{\mcitedefaultseppunct}\relax
\EndOfBibitem
\bibitem[Iwashina and Kudo(2011)Iwashina, and Kudo]{doi:10.1021/ja2050315}
Iwashina,~K.; Kudo,~A. Rh-doped SrTiO${}_{3}$ photocatalyst electrode showing
  cathodic photocurrent for water splitting under visible-light irradiation.
  \emph{Journal of the American Chemical Society} \textbf{2011}, \emph{133},
  13272--13275\relax
\mciteBstWouldAddEndPuncttrue
\mciteSetBstMidEndSepPunct{\mcitedefaultmidpunct}
{\mcitedefaultendpunct}{\mcitedefaultseppunct}\relax
\EndOfBibitem
\bibitem[Zhang \latin{et~al.}(2016)Zhang, Huang, Xu, Cao, Ho, and
  Lee]{doi:10.1021/acsami.5b11887}
Zhang,~Q.; Huang,~Y.; Xu,~L.; Cao,~J.-j.; Ho,~W.; Lee,~S.~C.
  Visible-light-active plasmonic Ag-SrTiO${}_{3}$ nanocomposites for the
  degradation of NO in air with high selectivity. \emph{ACS Applied Materials
  \& Interfaces} \textbf{2016}, \emph{8}, 4165--4174\relax
\mciteBstWouldAddEndPuncttrue
\mciteSetBstMidEndSepPunct{\mcitedefaultmidpunct}
{\mcitedefaultendpunct}{\mcitedefaultseppunct}\relax
\EndOfBibitem
\bibitem[Niishiro \latin{et~al.}(2014)Niishiro, Tanaka, and
  Kudo]{NIISHIRO2014187}
Niishiro,~R.; Tanaka,~S.; Kudo,~A. Hydrothermal-synthesized SrTiO${}_{3}$
  photocatalyst codoped with rhodium and antimony with visible-light response
  for sacrificial H${}_{2}$ and O${}_{2}$ evolution and application to overall
  water splitting. \emph{Applied Catalysis B: Environmental} \textbf{2014},
  \emph{150-151}, 187--196\relax
\mciteBstWouldAddEndPuncttrue
\mciteSetBstMidEndSepPunct{\mcitedefaultmidpunct}
{\mcitedefaultendpunct}{\mcitedefaultseppunct}\relax
\EndOfBibitem
\bibitem[Tan \latin{et~al.}(2014)Tan, Zhao, Zhu, Coker, Li, Zheng, Yu, Fan, and
  Sun]{doi:10.1021/am5051907}
Tan,~H.; Zhao,~Z.; Zhu,~W.-b.; Coker,~E.~N.; Li,~B.; Zheng,~M.; Yu,~W.;
  Fan,~H.; Sun,~Z. Oxygen vacancy enhanced photocatalytic activity of
  pervoskite SrTiO${}_{3}$. \emph{ACS Applied Materials \& Interfaces}
  \textbf{2014}, \emph{6}, 19184--19190\relax
\mciteBstWouldAddEndPuncttrue
\mciteSetBstMidEndSepPunct{\mcitedefaultmidpunct}
{\mcitedefaultendpunct}{\mcitedefaultseppunct}\relax
\EndOfBibitem
\bibitem[Zhang \latin{et~al.}(2017)Zhang, Wu, Tang, Li, Oropeza, Qiao, Lazarov,
  Du, Payne, MacManus-Driscoll, and Blamire]{doi:10.1021/acsami.7b06025}
Zhang,~K. H.~L.; Wu,~R.; Tang,~F.; Li,~W.; Oropeza,~F.~E.; Qiao,~L.;
  Lazarov,~V.~K.; Du,~Y.; Payne,~D.~J.; MacManus-Driscoll,~J.~L. \latin{et~al.}
   Electronic structure and band alignment at the NiO and SrTiO${}_{3}$ p-n
  heterojunctions. \emph{ACS Applied Materials \& Interfaces} \textbf{2017},
  \emph{9}, 26549--26555\relax
\mciteBstWouldAddEndPuncttrue
\mciteSetBstMidEndSepPunct{\mcitedefaultmidpunct}
{\mcitedefaultendpunct}{\mcitedefaultseppunct}\relax
\EndOfBibitem
\bibitem[Konta \latin{et~al.}(2004)Konta, Ishii, Kato, and
  Kudo]{doi:10.1021/jp049556p}
Konta,~R.; Ishii,~T.; Kato,~H.; Kudo,~A. Photocatalytic activities of noble
  metal ion doped SrTiO${}_{3}$ under visible light irradiation. \emph{The
  Journal of Physical Chemistry B} \textbf{2004}, \emph{108}, 8992--8995\relax
\mciteBstWouldAddEndPuncttrue
\mciteSetBstMidEndSepPunct{\mcitedefaultmidpunct}
{\mcitedefaultendpunct}{\mcitedefaultseppunct}\relax
\EndOfBibitem
\bibitem[Chen \latin{et~al.}(2012)Chen, Huang, Wu, and
  Lin]{doi:10.1021/jp300910e}
Chen,~H.-C.; Huang,~C.-W.; Wu,~J. C.~S.; Lin,~S.-T. Theoretical investigation
  of the metal-doped SrTiO${}_{3}$ photocatalysts for water splitting.
  \emph{The Journal of Physical Chemistry C} \textbf{2012}, \emph{116},
  7897--7903\relax
\mciteBstWouldAddEndPuncttrue
\mciteSetBstMidEndSepPunct{\mcitedefaultmidpunct}
{\mcitedefaultendpunct}{\mcitedefaultseppunct}\relax
\EndOfBibitem
\bibitem[Niishiro \latin{et~al.}(2005)Niishiro, Kato, and Kudo]{B502147B}
Niishiro,~R.; Kato,~H.; Kudo,~A. Nickel and either tantalum or niobium-codoped
  TiO${}_{2}$ and SrTiO${}_{3}$ photocatalysts with visible-light response for
  H${}_{2}$ or O${}_{2}$ evolution from aqueous solutions. \emph{Phys. Chem.
  Chem. Phys.} \textbf{2005}, \emph{7}, 2241--2245\relax
\mciteBstWouldAddEndPuncttrue
\mciteSetBstMidEndSepPunct{\mcitedefaultmidpunct}
{\mcitedefaultendpunct}{\mcitedefaultseppunct}\relax
\EndOfBibitem
\bibitem[Yu \latin{et~al.}(2011)Yu, Ouyang, Yan, Li, Yu, and Zou]{C1JM11385B}
Yu,~H.; Ouyang,~S.; Yan,~S.; Li,~Z.; Yu,~T.; Zou,~Z. Sol–gel hydrothermal
  synthesis of visible-light-driven Cr-doped SrTiO${}_{3}$ for efficient
  hydrogen production. \emph{J. Mater. Chem.} \textbf{2011}, \emph{21},
  11347--11351\relax
\mciteBstWouldAddEndPuncttrue
\mciteSetBstMidEndSepPunct{\mcitedefaultmidpunct}
{\mcitedefaultendpunct}{\mcitedefaultseppunct}\relax
\EndOfBibitem
\bibitem[Kawasaki \latin{et~al.}(2012)Kawasaki, Nakatsuji, Yoshinobu, Komori,
  Takahashi, Lippmaa, Mase, and Kudo]{doi:10.1063/1.4738371}
Kawasaki,~S.; Nakatsuji,~K.; Yoshinobu,~J.; Komori,~F.; Takahashi,~R.;
  Lippmaa,~M.; Mase,~K.; Kudo,~A. Epitaxial Rh-doped SrTiO${}_{3}$ thin film
  photocathode for water splitting under visible light irradiation.
  \emph{Applied Physics Letters} \textbf{2012}, \emph{101}, 033910\relax
\mciteBstWouldAddEndPuncttrue
\mciteSetBstMidEndSepPunct{\mcitedefaultmidpunct}
{\mcitedefaultendpunct}{\mcitedefaultseppunct}\relax
\EndOfBibitem
\bibitem[Liu \latin{et~al.}(2012)Liu, Nisar, Pathak, and Ahuja]{LIU201211611}
Liu,~P.; Nisar,~J.; Pathak,~B.; Ahuja,~R. Hybrid density functional study on
  SrTiO${}_{3}$ for visible light photocatalysis. \emph{International Journal
  of Hydrogen Energy} \textbf{2012}, \emph{37}, 11611--11617\relax
\mciteBstWouldAddEndPuncttrue
\mciteSetBstMidEndSepPunct{\mcitedefaultmidpunct}
{\mcitedefaultendpunct}{\mcitedefaultseppunct}\relax
\EndOfBibitem
\bibitem[Guo \latin{et~al.}(2015)Guo, Qiu, Dong, and Zhou]{C5CP03005F}
Guo,~Y.; Qiu,~X.; Dong,~H.; Zhou,~X. Trends in non-metal doping of the
  SrTiO${}_{3}$ surface: A hybrid density functional study. \emph{Phys. Chem.
  Chem. Phys.} \textbf{2015}, \emph{17}, 21611--21621\relax
\mciteBstWouldAddEndPuncttrue
\mciteSetBstMidEndSepPunct{\mcitedefaultmidpunct}
{\mcitedefaultendpunct}{\mcitedefaultseppunct}\relax
\EndOfBibitem
\bibitem[Wang \latin{et~al.}(2004)Wang, Yin, Komatsu, Zhang, Saito, and
  Sato]{WANG2004149}
Wang,~J.; Yin,~S.; Komatsu,~M.; Zhang,~Q.; Saito,~F.; Sato,~T. Preparation and
  characterization of nitrogen doped SrTiO${}_{3}$ photocatalyst. \emph{Journal
  of Photochemistry and Photobiology A: Chemistry} \textbf{2004}, \emph{165},
  149--156\relax
\mciteBstWouldAddEndPuncttrue
\mciteSetBstMidEndSepPunct{\mcitedefaultmidpunct}
{\mcitedefaultendpunct}{\mcitedefaultseppunct}\relax
\EndOfBibitem
\bibitem[Mi \latin{et~al.}(2006)Mi, Wang, Chai, Pan, Huan, Feng, and
  Ong]{doi:10.1063/1.2403181}
Mi,~Y.~Y.; Wang,~S.~J.; Chai,~J.~W.; Pan,~J.~S.; Huan,~C. H.~A.; Feng,~Y.~P.;
  Ong,~C.~K. Effect of nitrogen doping on optical properties and electronic
  structures of SrTiO${}_{3}$ films. \emph{Applied Physics Letters}
  \textbf{2006}, \emph{89}, 231922\relax
\mciteBstWouldAddEndPuncttrue
\mciteSetBstMidEndSepPunct{\mcitedefaultmidpunct}
{\mcitedefaultendpunct}{\mcitedefaultseppunct}\relax
\EndOfBibitem
\bibitem[Zhang \latin{et~al.}(2013)Zhang, Jia, Jing, Yao, Ma, and
  Sun]{ZHANG201369}
Zhang,~C.; Jia,~Y.; Jing,~Y.; Yao,~Y.; Ma,~J.; Sun,~J. Effect of non-metal
  elements (B, C, N, F, P, S) mono-doping as anions on electronic structure of
  SrTiO${}_{3}$. \emph{Computational Materials Science} \textbf{2013},
  \emph{79}, 69--74\relax
\mciteBstWouldAddEndPuncttrue
\mciteSetBstMidEndSepPunct{\mcitedefaultmidpunct}
{\mcitedefaultendpunct}{\mcitedefaultseppunct}\relax
\EndOfBibitem
\bibitem[Zou \latin{et~al.}(2012)Zou, Jiang, Qin, Zhao, Jiang, Zhi, Xiao, and
  Edwards]{C2CC33797E}
Zou,~F.; Jiang,~Z.; Qin,~X.; Zhao,~Y.; Jiang,~L.; Zhi,~J.; Xiao,~T.;
  Edwards,~P.~P. Template-free synthesis of mesoporous N-doped SrTiO${}_{3}$
  perovskite with high visible-light-driven photocatalytic activity.
  \emph{Chem. Commun.} \textbf{2012}, \emph{48}, 8514--8516\relax
\mciteBstWouldAddEndPuncttrue
\mciteSetBstMidEndSepPunct{\mcitedefaultmidpunct}
{\mcitedefaultendpunct}{\mcitedefaultseppunct}\relax
\EndOfBibitem
\bibitem[Reunchan \latin{et~al.}(2013)Reunchan, Ouyang, Umezawa, Xu, Zhang, and
  Ye]{C2TA00450J}
Reunchan,~P.; Ouyang,~S.; Umezawa,~N.; Xu,~H.; Zhang,~Y.; Ye,~J. Theoretical
  design of highly active SrTiO${}_{3}$-based photocatalysts by a codoping
  scheme towards solar energy utilization for hydrogen production. \emph{J.
  Mater. Chem. A} \textbf{2013}, \emph{1}, 4221--4227\relax
\mciteBstWouldAddEndPuncttrue
\mciteSetBstMidEndSepPunct{\mcitedefaultmidpunct}
{\mcitedefaultendpunct}{\mcitedefaultseppunct}\relax
\EndOfBibitem
\bibitem[Ouyang \latin{et~al.}(2012)Ouyang, Tong, Umezawa, Cao, Li, Bi, Zhang,
  and Ye]{doi:10.1021/ja210610h}
Ouyang,~S.; Tong,~H.; Umezawa,~N.; Cao,~J.; Li,~P.; Bi,~Y.; Zhang,~Y.; Ye,~J.
  Surface-alkalinization-induced enhancement of photocatalytic H${}_{2}$
  evolution over SrTiO${}_{3}$-based photocatalysts. \emph{Journal of the
  American Chemical Society} \textbf{2012}, \emph{134}, 1974--1977\relax
\mciteBstWouldAddEndPuncttrue
\mciteSetBstMidEndSepPunct{\mcitedefaultmidpunct}
{\mcitedefaultendpunct}{\mcitedefaultseppunct}\relax
\EndOfBibitem
\bibitem[Comes \latin{et~al.}(2014)Comes, Sushko, Heald, Colby, Bowden, and
  Chambers]{doi:10.1021/cm503541u}
Comes,~R.~B.; Sushko,~P.~V.; Heald,~S.~M.; Colby,~R.~J.; Bowden,~M.~E.;
  Chambers,~S.~A. Band-gap reduction and dopant interaction in epitaxial La,Cr
  co-doped SrTiO${}_{3}$ thin films. \emph{Chemistry of Materials}
  \textbf{2014}, \emph{26}, 7073--7082\relax
\mciteBstWouldAddEndPuncttrue
\mciteSetBstMidEndSepPunct{\mcitedefaultmidpunct}
{\mcitedefaultendpunct}{\mcitedefaultseppunct}\relax
\EndOfBibitem
\bibitem[Miyauchi \latin{et~al.}(2004)Miyauchi, Takashio, and
  Tobimatsu]{doi:10.1021/la0353125}
Miyauchi,~M.; Takashio,~M.; Tobimatsu,~H. Photocatalytic activity of
  SrTiO${}_{3}$ codoped with nitrogen and lanthanum under visible light
  illumination. \emph{Langmuir} \textbf{2004}, \emph{20}, 232--236\relax
\mciteBstWouldAddEndPuncttrue
\mciteSetBstMidEndSepPunct{\mcitedefaultmidpunct}
{\mcitedefaultendpunct}{\mcitedefaultseppunct}\relax
\EndOfBibitem
\bibitem[Chen \latin{et~al.}(2016)Chen, Liu, Li, Liu, Gao, Mao, Fan, Shangguan,
  Fang, and Liu]{CHEN2016145}
Chen,~W.; Liu,~H.; Li,~X.; Liu,~S.; Gao,~L.; Mao,~L.; Fan,~Z.; Shangguan,~W.;
  Fang,~W.; Liu,~Y. Polymerizable complex synthesis of SrTiO${}_{3}$: (Cr/Ta)
  photocatalysts to improve photocatalytic water splitting activity under
  visible light. \emph{Applied Catalysis B: Environmental} \textbf{2016},
  \emph{192}, 145--151\relax
\mciteBstWouldAddEndPuncttrue
\mciteSetBstMidEndSepPunct{\mcitedefaultmidpunct}
{\mcitedefaultendpunct}{\mcitedefaultseppunct}\relax
\EndOfBibitem
\bibitem[Wei \latin{et~al.}(2009)Wei, Dai, Guo, Yu, and
  Huang]{doi:10.1021/jp902567j}
Wei,~W.; Dai,~Y.; Guo,~M.; Yu,~L.; Huang,~B. Density functional
  characterization of the electronic structure and optical properties of
  N-Doped, La-Doped, and N/La-codoped SrTiO${}_{3}$. \emph{The Journal of
  Physical Chemistry C} \textbf{2009}, \emph{113}, 15046--15050\relax
\mciteBstWouldAddEndPuncttrue
\mciteSetBstMidEndSepPunct{\mcitedefaultmidpunct}
{\mcitedefaultendpunct}{\mcitedefaultseppunct}\relax
\EndOfBibitem
\bibitem[Wei \latin{et~al.}(2010)Wei, Dai, Guo, Yu, Jin, Han, and
  Huang]{B922399A}
Wei,~W.; Dai,~Y.; Guo,~M.; Yu,~L.; Jin,~H.; Han,~S.; Huang,~B. Codoping
  synergistic effects in N-doped SrTiO${}_{3}$ for higher energy conversion
  efficiency. \emph{Phys. Chem. Chem. Phys.} \textbf{2010}, \emph{12},
  7612--7619\relax
\mciteBstWouldAddEndPuncttrue
\mciteSetBstMidEndSepPunct{\mcitedefaultmidpunct}
{\mcitedefaultendpunct}{\mcitedefaultseppunct}\relax
\EndOfBibitem
\bibitem[Gai \latin{et~al.}(2009)Gai, Li, Li, Xia, and
  Wei]{PhysRevLett.102.036402}
Gai,~Y.; Li,~J.; Li,~S.-S.; Xia,~J.-B.; Wei,~S.-H. Design of narrow-gap
  ${\mathrm{TiO}}_{2}$: A passivated codoping approach for enhanced
  photoelectrochemical activity. \emph{Phys. Rev. Lett.} \textbf{2009},
  \emph{102}, 036402\relax
\mciteBstWouldAddEndPuncttrue
\mciteSetBstMidEndSepPunct{\mcitedefaultmidpunct}
{\mcitedefaultendpunct}{\mcitedefaultseppunct}\relax
\EndOfBibitem
\bibitem[Modak and Ghosh(2015)Modak, and Ghosh]{doi:10.1021/jp512948s}
Modak,~B.; Ghosh,~S.~K. Role of F in improving the photocatalytic activity of
  Rh-doped SrTiO${}_{3}$. \emph{The Journal of Physical Chemistry C}
  \textbf{2015}, \emph{119}, 7215--7224\relax
\mciteBstWouldAddEndPuncttrue
\mciteSetBstMidEndSepPunct{\mcitedefaultmidpunct}
{\mcitedefaultendpunct}{\mcitedefaultseppunct}\relax
\EndOfBibitem
\bibitem[Modak and Ghosh(2015)Modak, and Ghosh]{doi:10.1021/acs.jpcc.5b06667}
Modak,~B.; Ghosh,~S.~K. Enhancement of visible light photocatalytic activity of
  SrTiO${}_{3}$: A hybrid density functional study. \emph{The Journal of
  Physical Chemistry C} \textbf{2015}, \emph{119}, 23503--23514\relax
\mciteBstWouldAddEndPuncttrue
\mciteSetBstMidEndSepPunct{\mcitedefaultmidpunct}
{\mcitedefaultendpunct}{\mcitedefaultseppunct}\relax
\EndOfBibitem
\bibitem[Bahers and Takanabe(2019)Bahers, and Takanabe]{LEBAHERS2019212}
Bahers,~T.~L.; Takanabe,~K. Combined theoretical and experimental
  characterizations of semiconductors for photoelectrocatalytic applications.
  \emph{Journal of Photochemistry and Photobiology C: Photochemistry Reviews}
  \textbf{2019}, \emph{40}, 212--233\relax
\mciteBstWouldAddEndPuncttrue
\mciteSetBstMidEndSepPunct{\mcitedefaultmidpunct}
{\mcitedefaultendpunct}{\mcitedefaultseppunct}\relax
\EndOfBibitem
\bibitem[Walter \latin{et~al.}(2010)Walter, Warren, McKone, Boettcher, Mi,
  Santori, and Lewis]{doi:10.1021/cr1002326}
Walter,~M.~G.; Warren,~E.~L.; McKone,~J.~R.; Boettcher,~S.~W.; Mi,~Q.;
  Santori,~E.~A.; Lewis,~N.~S. Solar water splitting cells. \emph{Chemical
  Reviews} \textbf{2010}, \emph{110}, 6446--6473\relax
\mciteBstWouldAddEndPuncttrue
\mciteSetBstMidEndSepPunct{\mcitedefaultmidpunct}
{\mcitedefaultendpunct}{\mcitedefaultseppunct}\relax
\EndOfBibitem
\bibitem[Stevanovi\'c \latin{et~al.}(2014)Stevanovi\'c, Lany, Ginley, Tumas,
  and Zunger]{C3CP54589J}
Stevanovi\'c,~V.; Lany,~S.; Ginley,~D.~S.; Tumas,~W.; Zunger,~A. Assessing
  capability of semiconductors to split water using ionization potentials and
  electron affinities only. \emph{Phys. Chem. Chem. Phys.} \textbf{2014},
  \emph{16}, 3706--3714\relax
\mciteBstWouldAddEndPuncttrue
\mciteSetBstMidEndSepPunct{\mcitedefaultmidpunct}
{\mcitedefaultendpunct}{\mcitedefaultseppunct}\relax
\EndOfBibitem
\bibitem[Sun and Lu(2013)Sun, and Lu]{SUN2013176}
Sun,~T.; Lu,~M. Modification of SrTiO${}_{3}$ surface by nitrogen ion
  bombardment for enhanced photocatalysis. \emph{Applied Surface Science}
  \textbf{2013}, \emph{274}, 176--180\relax
\mciteBstWouldAddEndPuncttrue
\mciteSetBstMidEndSepPunct{\mcitedefaultmidpunct}
{\mcitedefaultendpunct}{\mcitedefaultseppunct}\relax
\EndOfBibitem
\bibitem[Liu \latin{et~al.}(2007)Liu, Zu, and Zhou]{Liu_2007}
Liu,~C.~M.; Zu,~X.~T.; Zhou,~W.~L. Photoluminescence of nitrogen doped
  {SrTiO}${}_{3}$. \emph{Journal of Physics D: Applied Physics} \textbf{2007},
  \emph{40}, 7318--7322\relax
\mciteBstWouldAddEndPuncttrue
\mciteSetBstMidEndSepPunct{\mcitedefaultmidpunct}
{\mcitedefaultendpunct}{\mcitedefaultseppunct}\relax
\EndOfBibitem
\bibitem[Tkach \latin{et~al.}(2005)Tkach, Vilarinho, and
  Kholkin]{TKACH20055061}
Tkach,~A.; Vilarinho,~P.~M.; Kholkin,~A.~L. Structure-microstructure-dielectric
  tunability relationship in Mn-doped strontium titanate ceramics. \emph{Acta
  Materialia} \textbf{2005}, \emph{53}, 5061--5069\relax
\mciteBstWouldAddEndPuncttrue
\mciteSetBstMidEndSepPunct{\mcitedefaultmidpunct}
{\mcitedefaultendpunct}{\mcitedefaultseppunct}\relax
\EndOfBibitem
\bibitem[Yang \latin{et~al.}(2014)Yang, Kotula, Sato, Chi, Ikuhara, and
  Browning]{doi:10.1080/21663831.2013.856815}
Yang,~H.; Kotula,~P.~G.; Sato,~Y.; Chi,~M.; Ikuhara,~Y.; Browning,~N.~D.
  Segregation of Mn${}^{2+}$ dopants as interstitials in SrTiO${}_{3}$ grain
  boundaries. \emph{Materials Research Letters} \textbf{2014}, \emph{2},
  16--22\relax
\mciteBstWouldAddEndPuncttrue
\mciteSetBstMidEndSepPunct{\mcitedefaultmidpunct}
{\mcitedefaultendpunct}{\mcitedefaultseppunct}\relax
\EndOfBibitem
\bibitem[Hohenberg and Kohn(1964)Hohenberg, and Kohn]{PhysRev.136.B864}
Hohenberg,~P.; Kohn,~W. Inhomogeneous electron gas. \emph{Phys. Rev.}
  \textbf{1964}, \emph{136}, B864--B871\relax
\mciteBstWouldAddEndPuncttrue
\mciteSetBstMidEndSepPunct{\mcitedefaultmidpunct}
{\mcitedefaultendpunct}{\mcitedefaultseppunct}\relax
\EndOfBibitem
\bibitem[Kohn and Sham(1965)Kohn, and Sham]{PhysRev.140.A1133}
Kohn,~W.; Sham,~L.~J. Self-consistent equations including exchange and
  correlation effects. \emph{Phys. Rev.} \textbf{1965}, \emph{140},
  A1133--A1138\relax
\mciteBstWouldAddEndPuncttrue
\mciteSetBstMidEndSepPunct{\mcitedefaultmidpunct}
{\mcitedefaultendpunct}{\mcitedefaultseppunct}\relax
\EndOfBibitem
\bibitem[Bhattacharya \latin{et~al.}(2017)Bhattacharya, Berger, Reuter,
  Ghiringhelli, and Levchenko]{PhysRevMaterials.1.071601}
Bhattacharya,~S.; Berger,~D.; Reuter,~K.; Ghiringhelli,~L.~M.; Levchenko,~S.~V.
  Theoretical evidence for unexpected O-rich phases at corners of MgO surfaces.
  \emph{Phys. Rev. Materials} \textbf{2017}, \emph{1}, 071601\relax
\mciteBstWouldAddEndPuncttrue
\mciteSetBstMidEndSepPunct{\mcitedefaultmidpunct}
{\mcitedefaultendpunct}{\mcitedefaultseppunct}\relax
\EndOfBibitem
\bibitem[Hedin(1965)]{PhysRev.139.A796}
Hedin,~L. New method for calculating the one-particle Green's function with
  application to the electron-gas problem. \emph{Phys. Rev.} \textbf{1965},
  \emph{139}, A796--A823\relax
\mciteBstWouldAddEndPuncttrue
\mciteSetBstMidEndSepPunct{\mcitedefaultmidpunct}
{\mcitedefaultendpunct}{\mcitedefaultseppunct}\relax
\EndOfBibitem
\bibitem[Hybertsen and Louie(1985)Hybertsen, and Louie]{PhysRevLett.55.1418}
Hybertsen,~M.~S.; Louie,~S.~G. First-principles theory of quasiparticles:
  Calculation of band gaps in semiconductors and insulators. \emph{Phys. Rev.
  Lett.} \textbf{1985}, \emph{55}, 1418--1421\relax
\mciteBstWouldAddEndPuncttrue
\mciteSetBstMidEndSepPunct{\mcitedefaultmidpunct}
{\mcitedefaultendpunct}{\mcitedefaultseppunct}\relax
\EndOfBibitem
\bibitem[Kresse and Furthm\"uller(1996)Kresse, and Furthm\"uller]{KRESSE199615}
Kresse,~G.; Furthm\"uller,~J. Efficiency of ab-initio total energy calculations
  for metals and semiconductors using a plane-wave basis set.
  \emph{Computational Materials Science} \textbf{1996}, \emph{6}, 15--50\relax
\mciteBstWouldAddEndPuncttrue
\mciteSetBstMidEndSepPunct{\mcitedefaultmidpunct}
{\mcitedefaultendpunct}{\mcitedefaultseppunct}\relax
\EndOfBibitem
\bibitem[Kresse and Joubert(1999)Kresse, and Joubert]{PhysRevB.59.1758}
Kresse,~G.; Joubert,~D. From ultrasoft pseudopotentials to the projector
  augmented-wave method. \emph{Phys. Rev. B} \textbf{1999}, \emph{59},
  1758--1775\relax
\mciteBstWouldAddEndPuncttrue
\mciteSetBstMidEndSepPunct{\mcitedefaultmidpunct}
{\mcitedefaultendpunct}{\mcitedefaultseppunct}\relax
\EndOfBibitem
\bibitem[Bl\"ochl(1994)]{PhysRevB.50.17953}
Bl\"ochl,~P.~E. Projector augmented-wave method. \emph{Phys. Rev. B}
  \textbf{1994}, \emph{50}, 17953--17979\relax
\mciteBstWouldAddEndPuncttrue
\mciteSetBstMidEndSepPunct{\mcitedefaultmidpunct}
{\mcitedefaultendpunct}{\mcitedefaultseppunct}\relax
\EndOfBibitem
\bibitem[Krukau \latin{et~al.}(2006)Krukau, Vydrov, Izmaylov, and
  Scuseria]{doi:10.1063/1.2404663}
Krukau,~A.~V.; Vydrov,~O.~A.; Izmaylov,~A.~F.; Scuseria,~G.~E. Influence of the
  exchange screening parameter on the performance of screened hybrid
  functionals. \emph{The Journal of Chemical Physics} \textbf{2006},
  \emph{125}, 224106\relax
\mciteBstWouldAddEndPuncttrue
\mciteSetBstMidEndSepPunct{\mcitedefaultmidpunct}
{\mcitedefaultendpunct}{\mcitedefaultseppunct}\relax
\EndOfBibitem
\bibitem[Monkhorst and Pack(1976)Monkhorst, and Pack]{PhysRevB.13.5188}
Monkhorst,~H.~J.; Pack,~J.~D. Special points for Brillouin-zone integrations.
  \emph{Phys. Rev. B} \textbf{1976}, \emph{13}, 5188--5192\relax
\mciteBstWouldAddEndPuncttrue
\mciteSetBstMidEndSepPunct{\mcitedefaultmidpunct}
{\mcitedefaultendpunct}{\mcitedefaultseppunct}\relax
\EndOfBibitem
\bibitem[Basera \latin{et~al.}(2019)Basera, Saini, Arora, Singh, Kumar, and
  Bhattacharya]{scireports2019}
Basera,~P.; Saini,~S.; Arora,~E.; Singh,~A.; Kumar,~M.; Bhattacharya,~S.
  Stability of non-metal dopants to tune the photo-absorption of TiO${}_{2}$ at
  realistic temperatures and oxygen partial pressures: A hybrid DFT study.
  \emph{Scientific reports} \textbf{2019}, \emph{9}, 1--13\relax
\mciteBstWouldAddEndPuncttrue
\mciteSetBstMidEndSepPunct{\mcitedefaultmidpunct}
{\mcitedefaultendpunct}{\mcitedefaultseppunct}\relax
\EndOfBibitem
\bibitem[Bhattacharya and Bhattacharya(2016)Bhattacharya, and
  Bhattacharya]{PhysRevB.94.094305}
Bhattacharya,~A.; Bhattacharya,~S. Unraveling the role of vacancies in the
  potentially promising thermoelectric clathrates
  $\textrm{Ba}_{8}\textrm{Zn}_{x}\textrm{Ge}_{46-x-y}\ensuremath{\square}_{y}$.
  \emph{Phys. Rev. B} \textbf{2016}, \emph{94}, 094305\relax
\mciteBstWouldAddEndPuncttrue
\mciteSetBstMidEndSepPunct{\mcitedefaultmidpunct}
{\mcitedefaultendpunct}{\mcitedefaultseppunct}\relax
\EndOfBibitem
\bibitem[Bhattacharya \latin{et~al.}(2014)Bhattacharya, Levchenko,
  Ghiringhelli, and Scheffler]{Bhattacharya_2014}
Bhattacharya,~S.; Levchenko,~S.~V.; Ghiringhelli,~L.~M.; Scheffler,~M.
  Efficient ab initio schemes for finding thermodynamically stable and
  metastable atomic structures: Benchmark of cascade genetic algorithms.
  \emph{New Journal of Physics} \textbf{2014}, \emph{16}, 123016\relax
\mciteBstWouldAddEndPuncttrue
\mciteSetBstMidEndSepPunct{\mcitedefaultmidpunct}
{\mcitedefaultendpunct}{\mcitedefaultseppunct}\relax
\EndOfBibitem
\bibitem[Bhattacharya and Bhattacharya(2015)Bhattacharya, and
  Bhattacharya]{doi:10.1021/acs.jpclett.5b01435}
Bhattacharya,~A.; Bhattacharya,~S. Exploring N-rich phases in Li$_x$N$_y$
  clusters for hydrogen storage at nanoscale. \emph{The Journal of Physical
  Chemistry Letters} \textbf{2015}, \emph{6}, 3726--3730\relax
\mciteBstWouldAddEndPuncttrue
\mciteSetBstMidEndSepPunct{\mcitedefaultmidpunct}
{\mcitedefaultendpunct}{\mcitedefaultseppunct}\relax
\EndOfBibitem
\bibitem[Arora \latin{et~al.}(2019)Arora, Saini, Basera, Kumar, Singh, and
  Bhattacharya]{ekta-jpcc}
Arora,~E.; Saini,~S.; Basera,~P.; Kumar,~M.; Singh,~A.; Bhattacharya,~S.
  Elucidating the role of temperature and pressure to the thermodynamic
  stability of charged defects in complex metal-hydrides: A case study of
  NaAlH${}_{4}$. \emph{The Journal of Physical Chemistry C} \textbf{2019},
  \emph{123}, 62--69\relax
\mciteBstWouldAddEndPuncttrue
\mciteSetBstMidEndSepPunct{\mcitedefaultmidpunct}
{\mcitedefaultendpunct}{\mcitedefaultseppunct}\relax
\EndOfBibitem
\bibitem[Li \latin{et~al.}(2006)Li, Wei, Li, and Xia]{PhysRevB.74.081201}
Li,~J.; Wei,~S.-H.; Li,~S.-S.; Xia,~J.-B. Design of shallow acceptors in
  $\mathrm{ZnO}$: First-principles band-structure calculations. \emph{Phys.
  Rev. B} \textbf{2006}, \emph{74}, 081201\relax
\mciteBstWouldAddEndPuncttrue
\mciteSetBstMidEndSepPunct{\mcitedefaultmidpunct}
{\mcitedefaultendpunct}{\mcitedefaultseppunct}\relax
\EndOfBibitem
\bibitem[Erg\"onenc \latin{et~al.}(2018)Erg\"onenc, Kim, Liu, Kresse, and
  Franchini]{PhysRevMaterials.2.024601}
Erg\"onenc,~Z.; Kim,~B.; Liu,~P.; Kresse,~G.; Franchini,~C. Converged $GW$
  quasiparticle energies for transition metal oxide perovskites. \emph{Phys.
  Rev. Materials} \textbf{2018}, \emph{2}, 024601\relax
\mciteBstWouldAddEndPuncttrue
\mciteSetBstMidEndSepPunct{\mcitedefaultmidpunct}
{\mcitedefaultendpunct}{\mcitedefaultseppunct}\relax
\EndOfBibitem
\bibitem[Cardona(1965)]{PhysRev.140.A651}
Cardona,~M. Optical properties and band structure of SrTi${\mathrm{O}}_{3}$ and
  BaTi${\mathrm{O}}_{3}$. \emph{Phys. Rev.} \textbf{1965}, \emph{140},
  A651--A655\relax
\mciteBstWouldAddEndPuncttrue
\mciteSetBstMidEndSepPunct{\mcitedefaultmidpunct}
{\mcitedefaultendpunct}{\mcitedefaultseppunct}\relax
\EndOfBibitem
\bibitem[Xu and Schoonen(2000)Xu, and Schoonen]{band_edge}
Xu,~Y.; Schoonen,~M. A.~A. The absolute energy positions of conduction and
  valence bands of selected semiconducting minerals. \emph{American
  Mineralogist} \textbf{2000}, \emph{85}, 543--556\relax
\mciteBstWouldAddEndPuncttrue
\mciteSetBstMidEndSepPunct{\mcitedefaultmidpunct}
{\mcitedefaultendpunct}{\mcitedefaultseppunct}\relax
\EndOfBibitem
\bibitem[Janotti \latin{et~al.}(2011)Janotti, Steiauf, and Van~de
  Walle]{PhysRevB.84.201304}
Janotti,~A.; Steiauf,~D.; Van~de Walle,~C.~G. Strain effects on the electronic
  structure of SrTiO${}_{3}$: Toward high electron mobilities. \emph{Phys. Rev.
  B} \textbf{2011}, \emph{84}, 201304\relax
\mciteBstWouldAddEndPuncttrue
\mciteSetBstMidEndSepPunct{\mcitedefaultmidpunct}
{\mcitedefaultendpunct}{\mcitedefaultseppunct}\relax
\EndOfBibitem
\bibitem[Marques \latin{et~al.}(2003)Marques, Teles, Anjos, Scolfaro, Leite,
  Freire, Farias, and da~Silva]{doi:10.1063/1.1570922}
Marques,~M.; Teles,~L.~K.; Anjos,~V.; Scolfaro,~L. M.~R.; Leite,~J.~R.;
  Freire,~V.~N.; Farias,~G.~A.; da~Silva,~E.~F. Full-relativistic calculations
  of the SrTiO${}_{3}$ carrier effective masses and complex dielectric
  function. \emph{Applied Physics Letters} \textbf{2003}, \emph{82},
  3074--3076\relax
\mciteBstWouldAddEndPuncttrue
\mciteSetBstMidEndSepPunct{\mcitedefaultmidpunct}
{\mcitedefaultendpunct}{\mcitedefaultseppunct}\relax
\EndOfBibitem
\bibitem[Fadlallah \latin{et~al.}(2018)Fadlallah, Shibl, Vlugt, and
  Schwingenschl\"ogl]{C8TA09022J}
Fadlallah,~M.; Shibl,~M.~F.; Vlugt,~T. J.~H.; Schwingenschl\"ogl,~U.
  Theoretical study on cation codoped SrTiO${}_{3}$ photocatalysts for water
  splitting. \emph{J. Mater. Chem. A} \textbf{2018}, \emph{6},
  24342--24349\relax
\mciteBstWouldAddEndPuncttrue
\mciteSetBstMidEndSepPunct{\mcitedefaultmidpunct}
{\mcitedefaultendpunct}{\mcitedefaultseppunct}\relax
\EndOfBibitem
\end{mcitethebibliography}


\providecommand{\latin}[1]{#1}
\makeatletter
\providecommand{\doi}
  {\begingroup\let\do\@makeother\dospecials
  \catcode`\{=1 \catcode`\}=2 \doi@aux}
\providecommand{\doi@aux}[1]{\endgroup\texttt{#1}}
\makeatother
\providecommand*\mcitethebibliography{\thebibliography}
\csname @ifundefined\endcsname{endmcitethebibliography}
  {\let\endmcitethebibliography\endthebibliography}{}
\begin{mcitethebibliography}{0}
\providecommand*\natexlab[1]{#1}
\providecommand*\mciteSetBstSublistMode[1]{}
\providecommand*\mciteSetBstMaxWidthForm[2]{}
\providecommand*\mciteBstWouldAddEndPuncttrue
  {\def\EndOfBibitem{\unskip.}}
\providecommand*\mciteBstWouldAddEndPunctfalse
  {\let\EndOfBibitem\relax}
\providecommand*\mciteSetBstMidEndSepPunct[3]{}
\providecommand*\mciteSetBstSublistLabelBeginEnd[3]{}
\providecommand*\EndOfBibitem{}
\mciteSetBstSublistMode{f}
\mciteSetBstMaxWidthForm{subitem}{(\alph{mcitesubitemcount})}
\mciteSetBstSublistLabelBeginEnd
  {\mcitemaxwidthsubitemform\space}
  {\relax}
  {\relax}

\end{mcitethebibliography}
\end{document}


\begin{center}
{\Large \bf Supplemental Material}\\ 
\end{center}
\begin{enumerate}[\bf I.]
	\item Band gap of different codopants in SrTiO$_3$ using PBE functional 
	\item Determination of chemical potential at different growth conditions.
	\item Optimized structure of defected and pristine SrTiO$_3$.
	\item Stability of defected SrTiO$_3$ at different environmental growth conditions
	\item Optical properties using HSE06 functional.
	\item Effect of functionals on phase diagrams.
	\item Dielectric constant ($\varepsilon_\infty$) using G$_0$W$_0$@HSE06
\end{enumerate}
\vspace*{12pt}
\clearpage
\section{Band gap of different codopants in SrTiO$_3$ using PBE functional}
Note that the choice of N/Mn is not fictitious. As a first step before choosing N/Mn, we did a careful prescreening of a large no. of various metals and nonmetals using PBE~\cite{PhysRevLett.77.3865} calculations. Even if PBE is not so good to estimate the band gap, it is good enough to give us some meaningful trend. The monodopants lead to the trap states that are detrimental to the photocatalysis. This has motivated us, to try different codopants and we have identified some of them (see the Table~\ref{tbl:ex} and the no.s marked as red) to be promising for photocatalysis, where the band gap for maximum efficiency should be $\sim$ 2 eV~\cite{doi:10.1021/cr1002326,C3CP54589J}. Note that in this class of oxide perovskite, the typical error in PBE band gap with respect to HSE06~\cite{hybrid_2006} is $\sim$ 1.0-1.5 eV [for example, the pristine has the PBE (HSE06) band gap of 1.75 (3.28) eV]. Therefore, keeping this in mind we have identified few promising systems (marked as red) and one of them is the N and Mn codoped system. Incidentally, we have also noted that the individual monodopants (i.e. N and Mn) are already experimentally synthesized~\cite{doi:10.1063/1.2403181,SUN2013176,Liu_2007,TKACH20055061,doi:10.1080/21663831.2013.856815}. Hence, we have proposed, presumably for the first time, N and Mn as codopants. The band gap of all the considered codoped cases are given below in Table~\ref{tbl:ex}.
\begin{table}[H]
	\caption{Band gap of different codopants in SrTiO$_3$ using PBE functional}
	\label{tbl:ex}
	\begin{tabular}{|c|c|}
		\hline
		Codoped System  & Band gap (eV)  \\
		\hline
		Cr$_\textrm{Ti}$B$_\textrm{O}$   & \textcolor{red}{0.57}   \\
		Cr$_\textrm{Ti}$C$_\textrm{O}$ & metallic  \\
		Cr$_\textrm{Ti}$F$_\textrm{O}$  & 1.13  \\
		Cr$_\textrm{Ti}$N$_\textrm{O}$ & 0.34 \\
		Cr$_\textrm{Ti}$S$_\textrm{O}$ & 0.35 \\
		Fe$_\textrm{Ti}$B$_\textrm{O}$   & 0.17   \\
		Fe$_\textrm{Ti}$C$_\textrm{O}$ & metallic  \\
		Fe$_\textrm{Ti}$F$_\textrm{O}$  & 0.30  \\
		Fe$_\textrm{Ti}$N$_\textrm{O}$ & metallic \\
		Fe$_\textrm{Ti}$S$_\textrm{O}$ & metallic \\
		Mn$_\textrm{Ti}$B$_\textrm{O}$   & 0.38   \\
		Mn$_\textrm{Ti}$C$_\textrm{O}$ & 0.13 \\
		Mn$_\textrm{Ti}$F$_\textrm{O}$  & 1.39  \\
		Mn$_\textrm{Ti}$N$_\textrm{O}$ & \textcolor{red}{0.70} \\
		Mn$_\textrm{Ti}$S$_\textrm{O}$ & \textcolor{red}{0.89} \\
		Rh$_\textrm{Ti}$B$_\textrm{O}$   & 0.07   \\
		Rh$_\textrm{Ti}$C$_\textrm{O}$ & 0.34 \\
		Rh$_\textrm{Ti}$F$_\textrm{O}$  & \textcolor{red}{1.03}  \\
		Rh$_\textrm{Ti}$N$_\textrm{O}$ & 1.19 \\
		Rh$_\textrm{Ti}$S$_\textrm{O}$ & \textcolor{red}{0.85} \\
		La$_\textrm{Sr}$B$_\textrm{O}$   & 1.41   \\
		La$_\textrm{Sr}$C$_\textrm{O}$ & 0.27 \\
		La$_\textrm{Sr}$F$_\textrm{O}$  & 2.19  \\
		La$_\textrm{Sr}$N$_\textrm{O}$ & 1.42 \\
		La$_\textrm{Sr}$S$_\textrm{O}$ & 1.43 \\
		Pr$_\textrm{Sr}$B$_\textrm{O}$   & 1.33   \\
		Pr$_\textrm{Sr}$C$_\textrm{O}$ & 1.65 \\
		Pr$_\textrm{Sr}$F$_\textrm{O}$  & 2.06  \\
		Pr$_\textrm{Sr}$N$_\textrm{O}$ & 1.25 \\
		Pr$_\textrm{Sr}$S$_\textrm{O}$ & 1.47 \\
		\hline
	\end{tabular}
\end{table}
\section{Determination of chemical potential at different growth conditions}
The effect of temperature and pressure is explicitly taken into chemical potential term. The chemical potential $\Delta\mu_\textrm{O}$ as a function of temperature ($T$) and the partial pressure of oxygen ($p_{\textrm{O}_2}$) is calculated using the relation~\cite{Bhattacharya_2014}:
\begin{equation}\begin{split}
\Delta\mu_\textrm{O}(T, p_{\textrm{O}_2}) &= \frac{1}{2}\left[ -k_\textrm{B}T \ln\left[\left(\frac{2\pi m}{h^2}\right)^\frac{3}{2}\left(k_\textrm{B}T\right)^\frac{5}{2}\right]\right.
\\
&\quad+ k_\textrm{B}T \ln p_{\textrm{O}_2} - k_\textrm{B}T \ln \left(\frac{8\pi^2I_Ak_\textrm{B}T}{h^2}\right)
\\
&\quad+ k_\textrm{B}T \ln \left[1-\exp\left(\frac{-h\nu_\textrm{OO}}{k_\textrm{B}T}\right)\right]
\\
&\quad\left.-  k_\textrm{B}T \ln \mathcal{M} + k_\textrm{B}T \ln \sigma \right]\textrm{,}
\end{split}\end{equation}
where $m$ is the mass, $I_A$ is the moment of inertia of $\textrm{O}_2$ molecule, $\nu_\textrm{OO}$ is the O-O stretching frequency, $\mathcal{M}$ is the spin multiplicity and $\sigma$ is the symmetry number.

Under equilibrium growth conditions, the chemical potentials are related to enthalpy of formation of SrTiO$_3$ $(\Delta\textrm{H}_\textrm{f}(\textrm{SrTiO}_3))$ by:
\begin{equation}
\Delta\mu_\textrm{Sr} + \Delta\mu_\textrm{Ti} + 3\Delta\mu_\textrm{O} = \Delta\textrm{H}_\textrm{f}(\textrm{SrTiO}_3)\textrm{.}
\end{equation}
To ensure the suppression of secondary phases, constraints are imposed on the different chemical potentials as given below:
\begin{equation}\begin{split}
\Delta\mu_\textrm{Ti} + 2\Delta\mu_\textrm{O} \leq \Delta\textrm{H}_\textrm{f}(\textrm{TiO}_2)
\\
\Delta\mu_\textrm{Sr} + \Delta\mu_\textrm{O} \leq \Delta\textrm{H}_\textrm{f}(\textrm{SrO})
\\
\Delta\mu_\textrm{X} \leq 0\textrm{.}
\end{split}\end{equation}
\begin{table}
	\caption{The chemical potentials at different environmental conditions}
	\label{tbl:example}
	\begin{tabular}{llllll}
		\hline
		Growth Conditions & $\;\;\Delta\mu_\textrm{O}$ & $\;\;\Delta\mu_\textrm{N}$ & $\;\;\Delta\mu_\textrm{Ti}$ & $\;\;\Delta\mu_\textrm{Sr}$ & $\,\Delta\mu_\textrm{Mn}$ \\
		\hline
		O-poor (Ti-rich) & $-4.55$ & $-1.48$  & $\;\;\:0.00$ & $-3.98$ & $\;\;\:0.00$  \\
		O-intermediate & $-1.58$ & $-1.48$ & $-5.95$ & $-6.96$  & $-2.44$  \\
		O-rich (Ti-poor) & $\;\;\:0.00$ & $-1.48$ & $-9.11$ & $-8.54$ & $-5.61$  \\
		\hline
	\end{tabular}
\end{table}
We have calculated the formation energy in three regimes, viz. O-rich, O-intermediate and O-poor conditions. The $\Delta\mu_\textrm{X}$ for different growth conditions are given in Table~\ref{tbl:example}. Under O-rich and O-intermediate conditions, $\Delta\mu_\textrm{Ti}$ and $\Delta\mu_\textrm{Mn}$ are limited by the formation of TiO$_2$ and MnO$_2$, respectively. In O-intermediate condition, $\Delta\mu_\textrm{O}$ takes the value to reflect the experimental growth condition ($T$ = 1100 $\degree\textrm{C}$, $p_{\textrm{O}_2} = 1$ atm~\cite{WANG2004149}). Under O-poor (Ti-rich) condition, $\Delta\mu_\textrm{Ti}$ and $\Delta\mu_\textrm{Mn}$ are limited by the formation of metallic phase of Ti and Mn, respectively. We have fixed the $\Delta\mu_\textrm{N} = -1.48$ eV, which is obtained at experimental growth condition. By knowing the stability of different dopants under certain conditions, one could deliberately dope the SrTiO$_3$ in accordance with the need.
\section{Optimized structure of defected and pristine SrTiO$_3$}
\begin{figure}[H]
	\centering
	\includegraphics[width=0.6\textwidth]{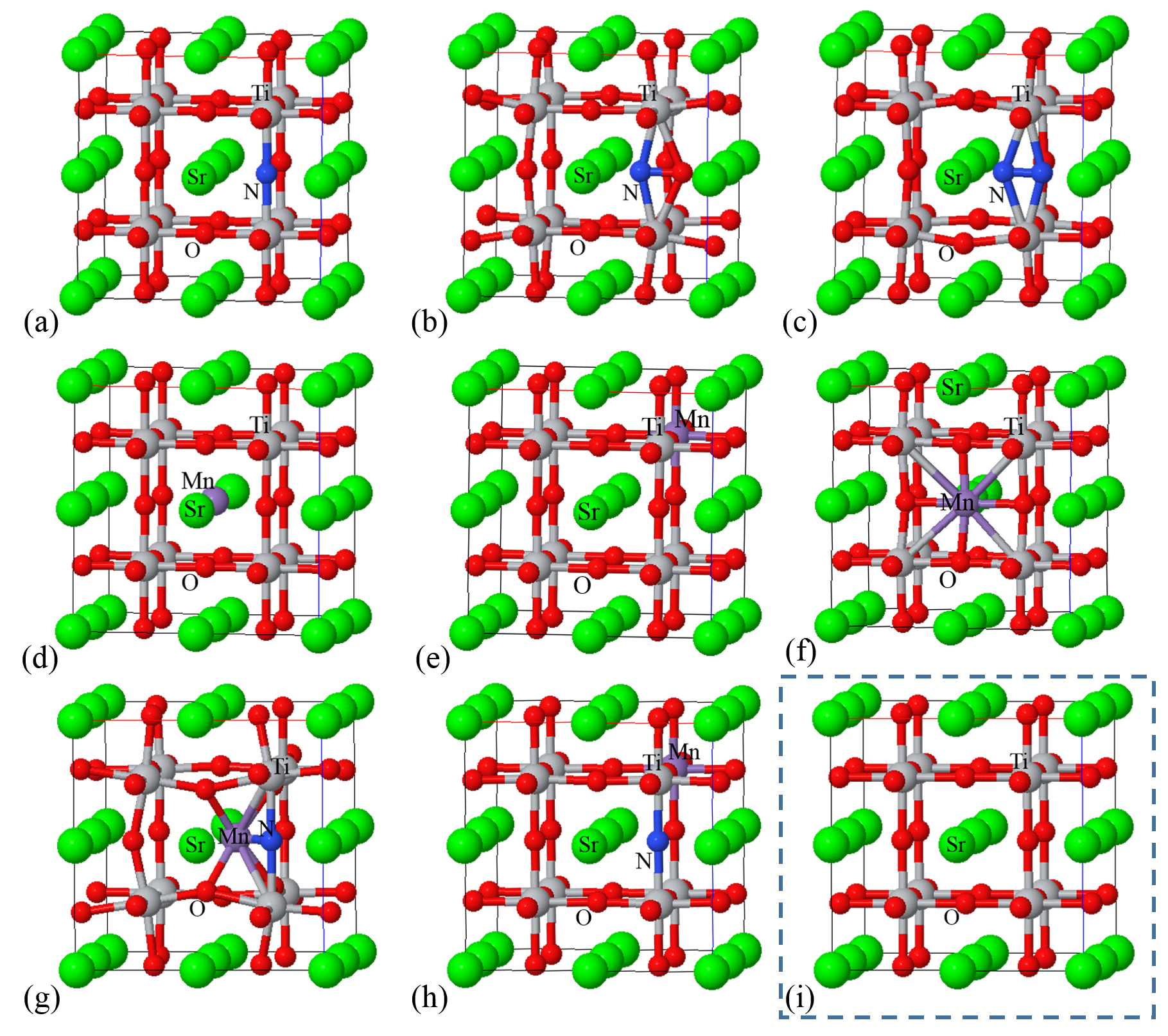} 
	\caption{Ball and stick model of optimized structures of (a) N$_\textrm{O}$, (b) N$_\textrm{i}$, (c) $(\textrm{N}_2)_\textrm{O}$, (d) Mn$_\textrm{Sr}$, (e) Mn$_\textrm{Ti}$, (f) Mn$_\textrm{i}$ (g) $\textrm{Mn}_\textrm{Sr}\textrm{N}_\textrm{O}$, (h) $\textrm{Mn}_\textrm{Ti}\textrm{N}_\textrm{O}$, and (i) pristine SrTiO$_3$.}
	\label{fig_struct}
\end{figure}
\section{Stability of defected SrTiO$_3$ at different environmental growth conditions}
\begin{figure}[H]
	\centering
	\includegraphics[width=0.8\textwidth]{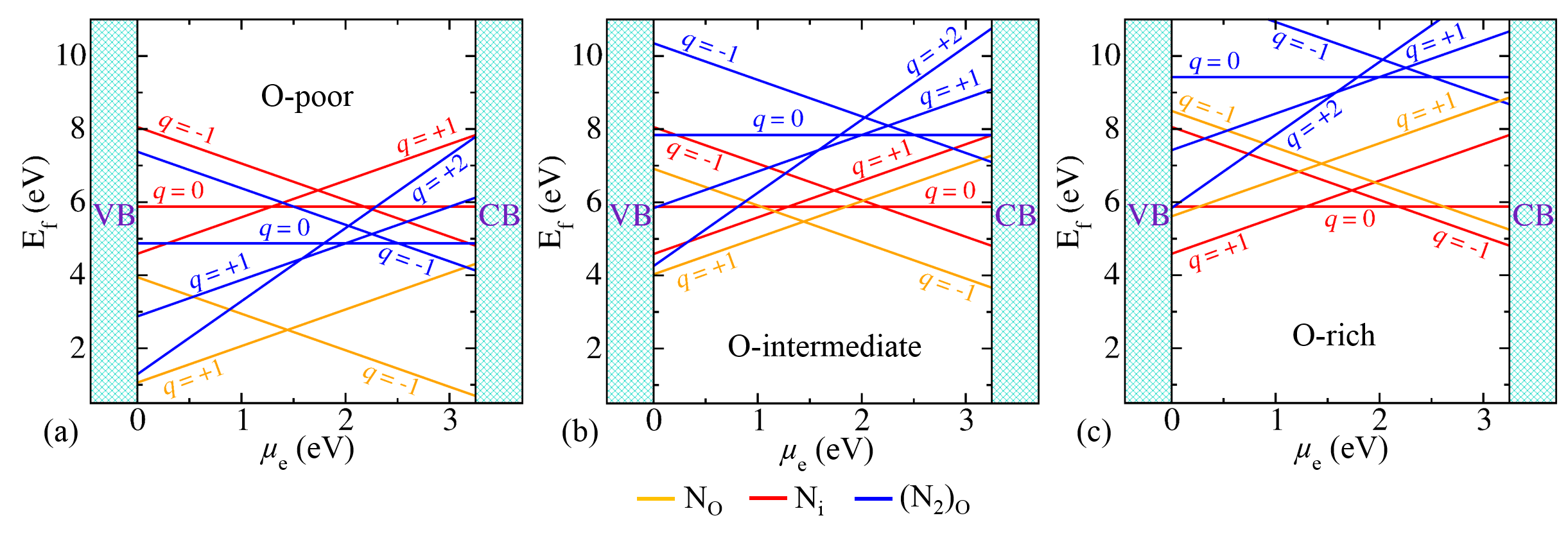}
	\caption{Formation energy of N-related defects as a function of chemical potential of electron under (a) O-poor, (b) O-intermediate and (c) O-rich condition.}
	\label{fig_N-rel}
\end{figure}

\begin{figure}[H]
	\centering
	\includegraphics[width=0.8\textwidth]{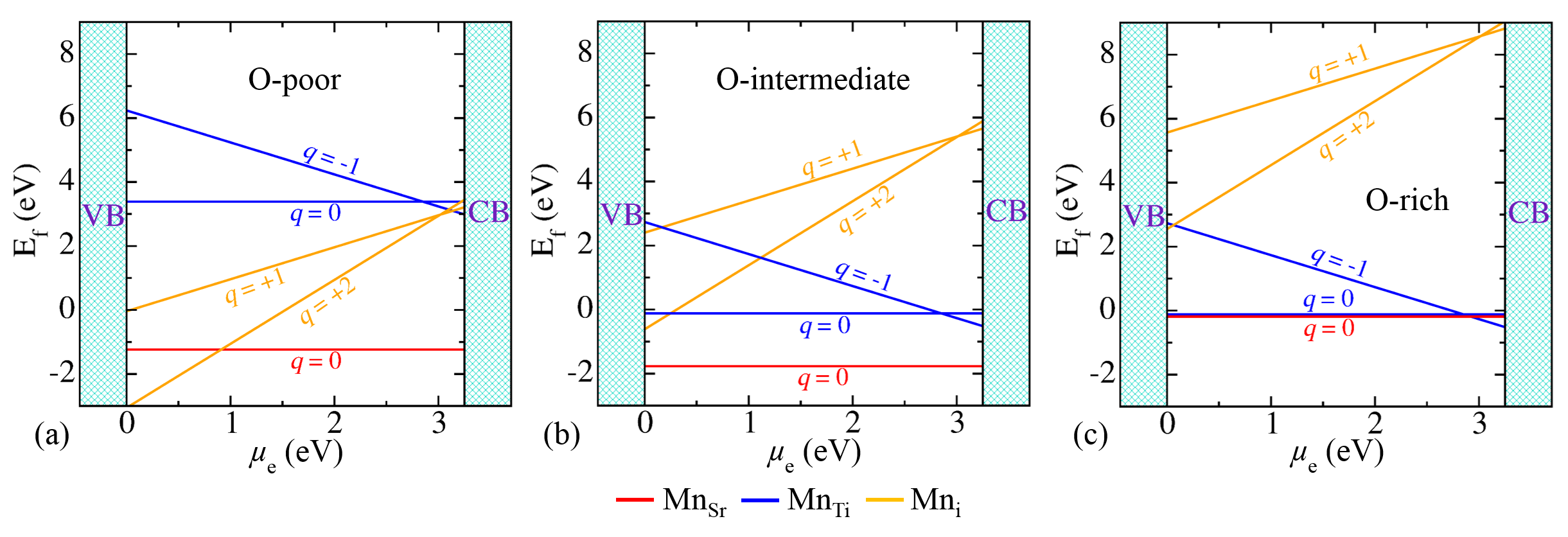}
	\caption{Formation energy of Mn-related defects as a function of chemical potential of electron under (a) O-poor, (b) O-intermediate and (c) O-rich condition.}
	\label{fig_Mn-rel}
\end{figure}

\begin{figure}[H]
	\centering
	\includegraphics[width=0.8\textwidth]{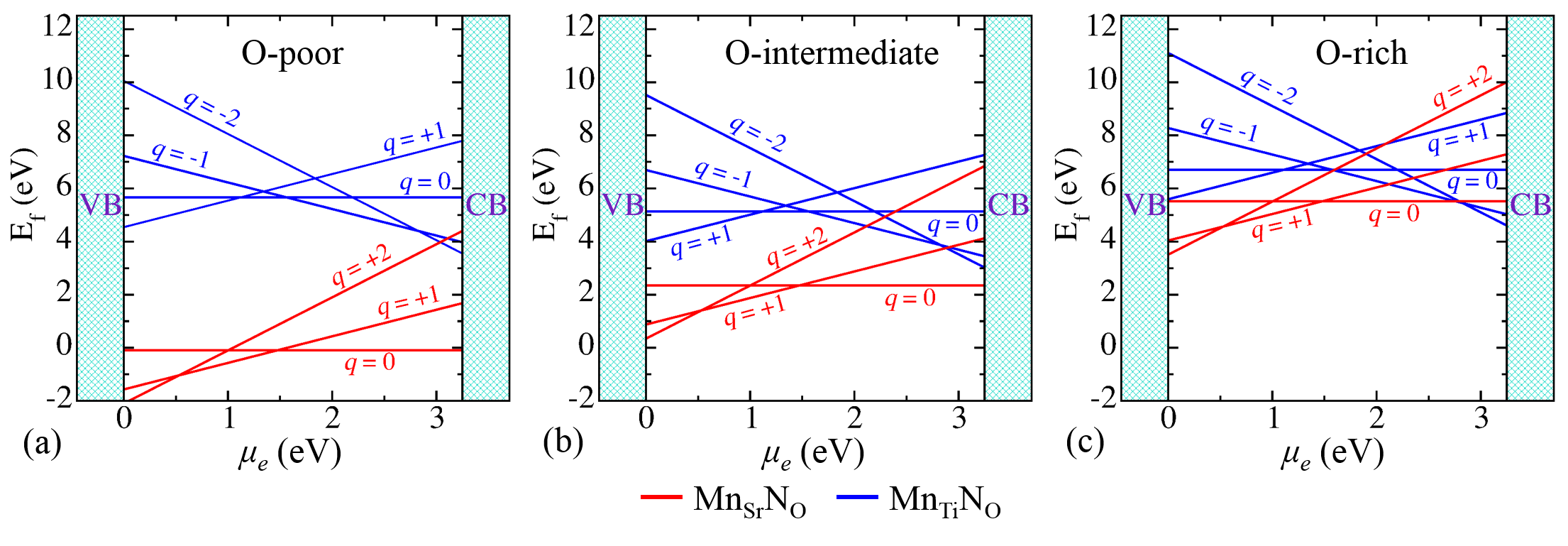}
	\caption{Formation energy of (Mn+N)-related defects as a function of chemical potential of electron under (a) O-poor, (b) O-intermediate and (c) O-rich condition.}
	\label{fig_Mn-N-rel}
\end{figure}

The formation energies of N-, Mn- and (N-Mn)-related defects as a function of $\mu_\textrm{e}$ with different charge states, under different environmental conditions are shown in Figure~\ref{fig_N-rel},~\ref{fig_Mn-rel} and~\ref{fig_Mn-N-rel}, respectively. Only those charge states of a particular defect are shown that have lowest formation energies.
\section{Optical properties using HSE06 functional}
The real part $\epsilon_1(\omega)$ is calculated using the Kramers-Kronig transformation, and the imaginary part $\epsilon_2(\omega)$ is evaluated by taking summation over large number of empty states as implemented in VASP. The absorption coefficient $\alpha(\omega)$ is related to real and imaginary part of dielectric function as follow:
\begin{equation}
\alpha(\omega) = \sqrt{2}\,\omega\left(\sqrt{\epsilon_1(\omega)^2 + \epsilon_2(\omega)^2} - \epsilon_1(\omega)\right)^{\frac{1}{2}}\textrm{.}
\end{equation}
\begin{figure}[H]
	\centering
	\includegraphics[width=0.25\columnwidth,clip]{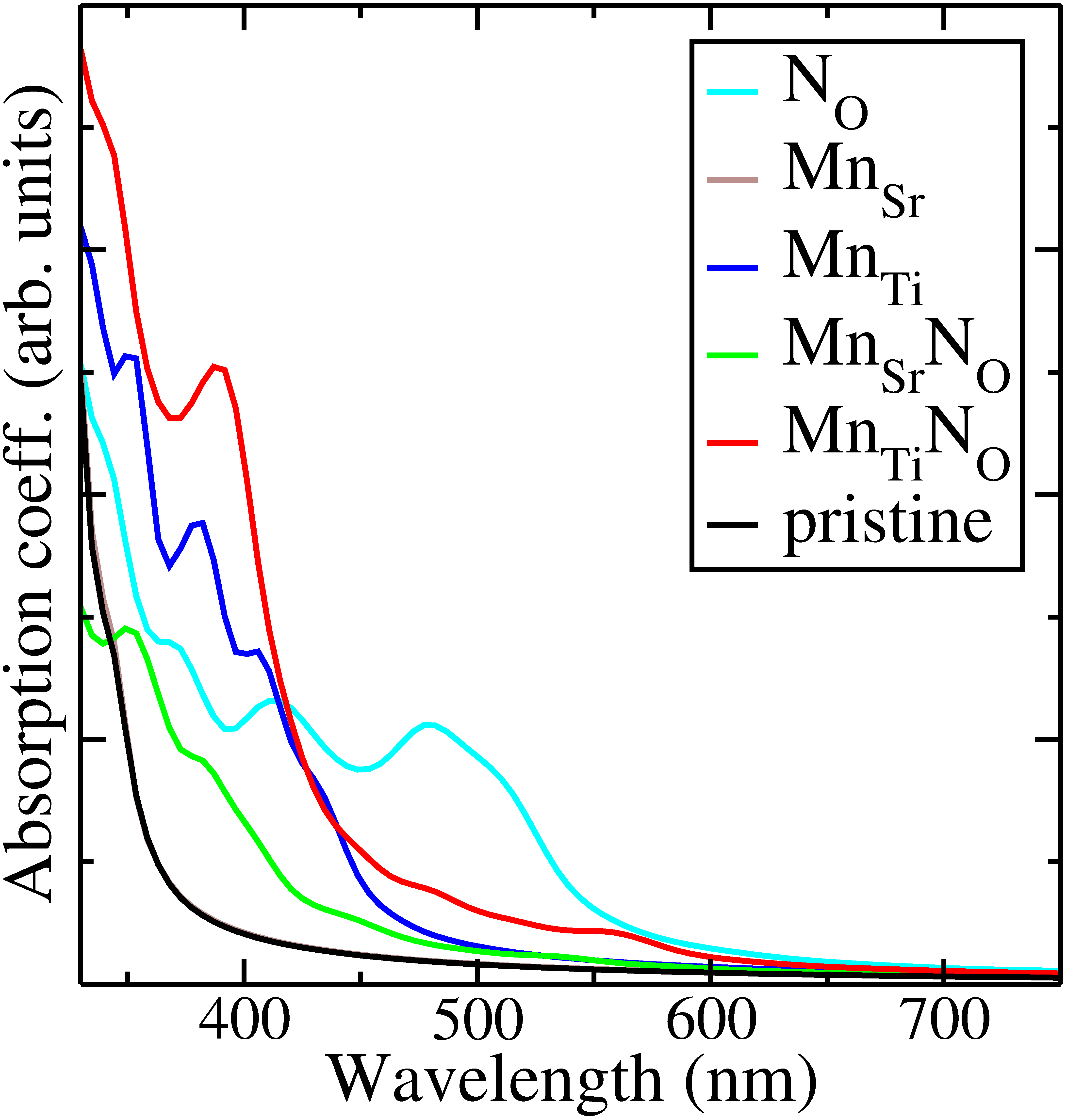}
	\caption{Absorption plot for undoped, monodoped and codoped SrTiO$_3$ using HSE06 functional.}
	\label{fig_abs}
\end{figure}
\begin{figure}[H]
	\centering
	\includegraphics[width=0.5\columnwidth,clip]{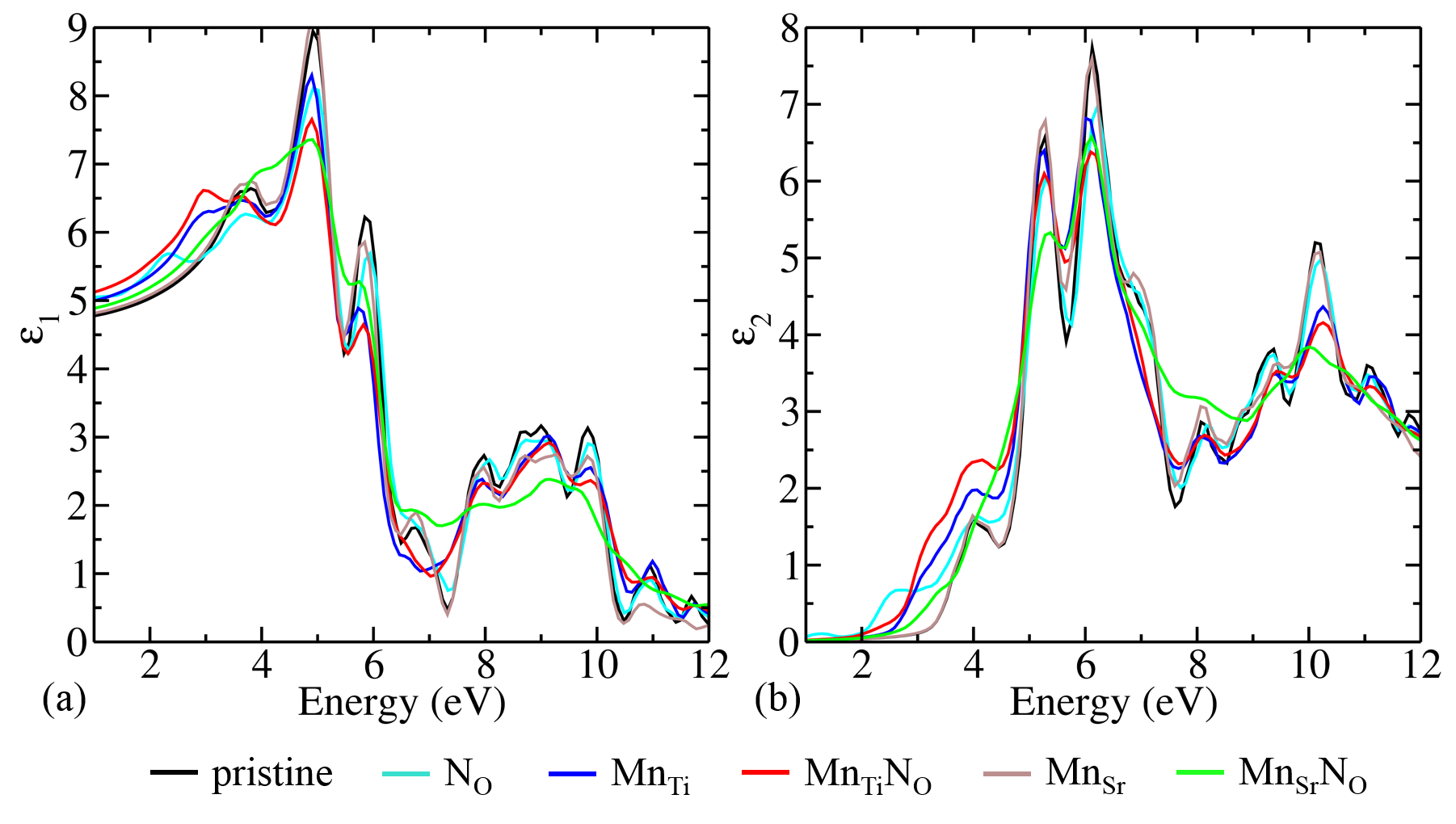}
	\caption{Spatially average (a) real $(\epsilon_1)$ and (b) imaginary $(\epsilon_2)$ part of the dielectric function obtained by HSE06 for the pristine, monodoped and codoped SrTiO$_3$.}
	\label{fig_opt}
\end{figure}
\section{Effect of functionals on phase diagrams}
We have benchmarked the exchange-correlation (xc) functionals viz. local-density approximation (LDA~\cite{perdew_wang_prb1992}), semi-local generalized gradient approximation (PBE) and a more pronounced non-local hybrid functional HSE06 to ensure that our results are not an artifact of chosen treatment for the xc. The LDA and PBE functionals underestimate the band gap, giving a value of 1.37 eV and 1.75 eV, respectively. While, the hybrid functional HSE06 reproduce the band gap of 3.28 eV by taking the 28\% of Hartree-Fock exact exchange into account, which is in nice match with the experimental value of 3.25 eV~\cite{expt_gap}. We have also calculated the defect formation energy for single oxygen vacancy E$_\textrm{f}(\framebox(6,6){})^q$ in our system with charge $q = 0, +1, +2, -1, \textrm{and} -2$ using LDA, PBE and HSE06 functionals, since we have to find out the stability of defects as well as the other energetics. The E$_\textrm{f}(\framebox(6,6){})^q$ has been calculated as follow~\cite{scireports2019,PhysRevB.94.094305,sasprmr17}:
\begin{equation}\begin{split}
\textrm{E}_\textrm{f}(\framebox(6,6){})^{q} &= \textrm{E}_\textrm{tot}(\framebox(6,6){})^{q} - \textrm{E}_\textrm{tot}(\textrm{SrTiO}_3) + \mu_\textrm{O}
\\
&\quad+ {q}(\mu_\textrm{e} + \textrm{VBM} + \Delta\textrm{V})\textrm{,}
\end{split}\end{equation}
where $\textrm{E}_\textrm{tot}(\framebox(6,6){})^{q}$ is the total energy of supercell containing single oxygen vacancy with charge state $q$, and $\textrm{E}_\textrm{tot}(\textrm{SrTiO}_3)$ is the total energy of the same supercell without any defect. $\mu_\textrm{O}$ is the chemical potential of the oxygen atom, which is equal to the energy required to remove one O
atom from the pristine supercell and put it into the O
reservoir; $\mu_\textrm{O} = \frac{1}{2}\textrm{E}_\textrm{tot}(\textrm{O}_2)$, and $\textrm{E}_\textrm{tot}(\textrm{O}_2)$ is the total energy of isolated O$_2$ molecule. $\mu_\textrm{e}$ is the chemical potential of electron, that is referenced from VBM of the pristine supercell. $\Delta\textrm{V}$ is the core level alignment between the supercells with and without defect.\\
\begin{figure}[H]
	\centering
	\includegraphics[width=0.5\columnwidth,clip]{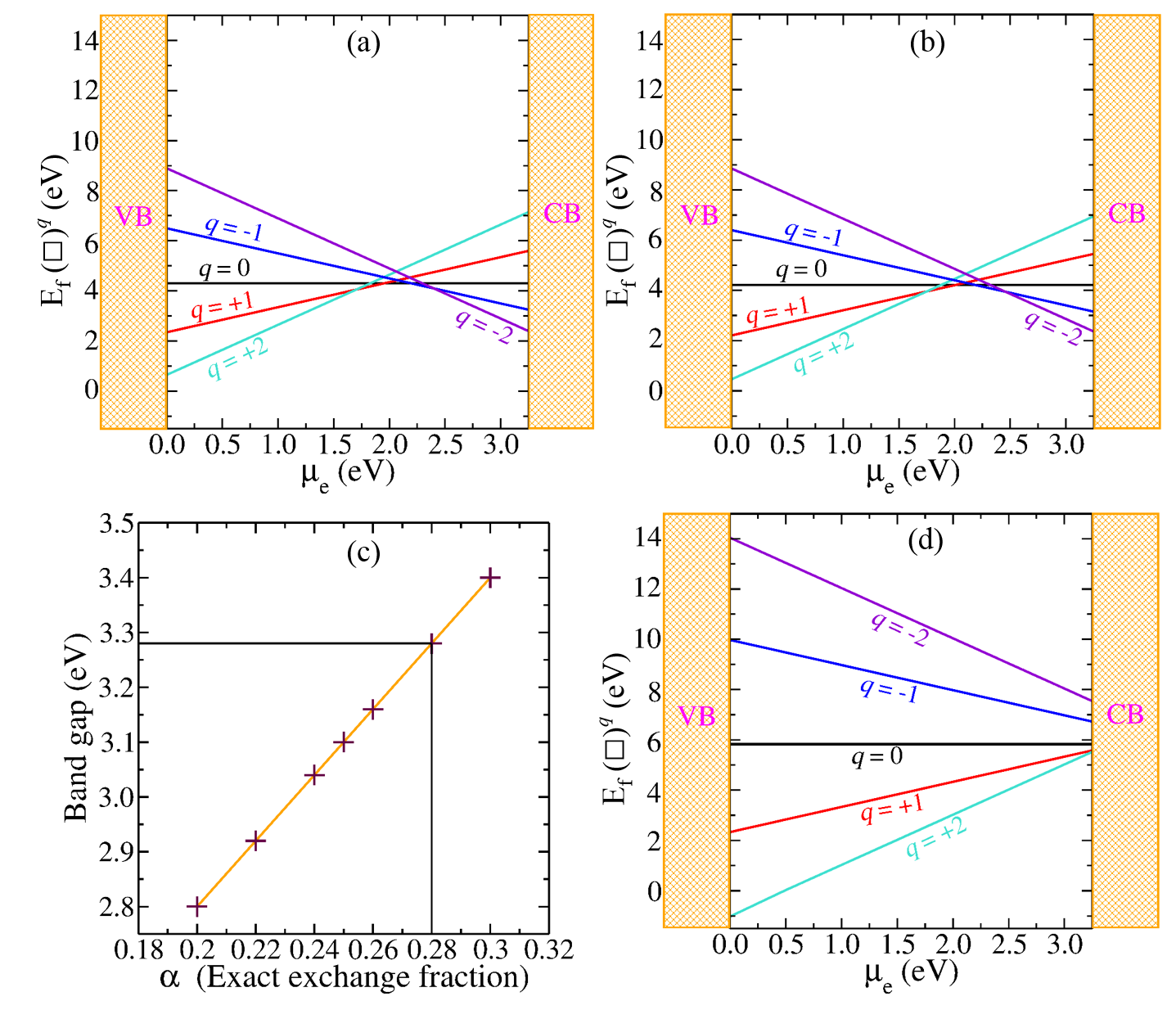}
	\caption{Formation energy of single oxygen vacancy defect as a function of chemical potential of electron under O-rich condition using (a) LDA, (b) PBE, and (d) HSE06 xc functional. (c) represents the variation of bandgap of pristine supercell as a function of exact exchange fraction contained in HSE06 functional.}
	\label{fig_val}
\end{figure}
We have found from Figure~\ref{fig_val} that $q = +2$ and $q = -2$ are the stable charge states near VBM and CBm respectively using both LDA and PBE functionals. However, this is not the case with hybrid functional HSE06. With HSE06, we have observed that only $q = +2$ charge state is stable througout the band gap. It implies that, results which are obtained from LDA, PBE xc functionals are different from the HSE06 functional. And as a matter of fact, HSE06 functional is more accurate being non-local and containing a fraction of exact exchange, HSE06 xc functional is more reliable for our system. Therefore, we have done our further calculations with HSE06 functional. Note that we have used a 40-atom supercell to introduce defects in SrTiO$_3$, which is a $2\,\times\,2\,\times\,2$ repetition of cubic SrTiO$_3$ unit cell (5 atoms). To ensure the convergence of supercell size, test calculations have been performed with 90-atom supercell ($3\times3\times2$ repetition of unit cell of SrTiO$_3$) for the case of N$_\textrm{O}$ in order to ensure that the defect is fully localized. The results obtained from 40- and 90-atom supercells are consistent with each other. Therefore, we have performed all the calculations with 40-atom supercell.
\section{Dielectric constant ($\varepsilon_\infty$) using G$_0$W$_0$@HSE06}
\begin{table}
	\caption{The high frequency `ion-clamped' dielectric constant ($\varepsilon_\infty$)}
	\label{tbl}
	\begin{tabular}{ll}
		\hline
		Configurations & $\varepsilon_\infty$ \\
		\hline
		Pristine & $3.46$\\
		N$_\textrm{O}$ & $3.61$\\
		Mn$_\textrm{Sr}$ & $3.48$\\
		Mn$_\textrm{Ti}$ & $3.66$\\
		Mn$_\textrm{Sr}$N$_\textrm{O}$ & $3.54$\\
		Mn$_\textrm{Ti}$N$_\textrm{O}$ & $3.72$\\
		\hline
	\end{tabular}
\end{table}
{}